\documentclass[useAMS,usenatbib]{mn2e}
\usepackage{epsfig}
\usepackage{amsmath}
\usepackage{rotating}
\usepackage{multirow}
\usepackage{natbib}

\newcommand{\be}{\begin{equation}}
\newcommand{\beq}{\begin{equation}}
\newcommand{\ba}{\begin{eqnarray}}
\newcommand{\ee}{\end{equation}}
\newcommand{\eeq}{\end{equation}}
\newcommand{\ea}{\end{eqnarray}}
\newcommand{\msun}{\rm M_{\odot}}

\def\lsim{~\rlap{$<$}{\lower 1.0ex\hbox{$\sim$}}}

\def\gsim{~\rlap{$>$}{\lower 1.0ex\hbox{$\sim$}}}

\title[Improved Models for CIB Anisotropy Power Spectrum]{Improved
Models for Cosmic Infrared Background Anisotropies: New Constraints on
the IR Galaxy Population}
\author[Shang et al.]{Cien Shang$^{1}$, Zolt\'an Haiman$^{2}$, Lloyd Knox$^{3}$, S. Peng Oh$^{4}$\\
$^{1}$Kavli Insitute for Theoretical Physics, University of California, Santa Barbara, California, CA 93106; cshang@kitp.ucsb.edu\\
$^{2}$Department of Astronomy, Columbia University, 550 West 120th Street, New York, NY 10027, USA; zoltan@astro.columbia.edu\\
$^{3}$Deparment of Physics, University of California, Davis, CA 95616; lknox@ucdavis.edu\\
$^{4}$Department of Physics, University of California, Santa Barbara, California, CA 93106; peng@physics.ucsb.edu}
\begin{document}
\bibliographystyle{mn2e}

\date{\today}
\pagerange{\pageref{firstpage}--\pageref{lastpage}} \pubyear{2009}

\maketitle

\label{firstpage}

\begin{abstract}
The power spectrum of cosmic infrared background (CIB) anisotropies is
sensitive to the connection between star formation and dark matter
halos over the entire cosmic star formation history.  Here we develop
a model that associates star-forming galaxies with dark matter halos
and their subhalos.  The model is based on a parameterized relation
between the dust-processed infrared luminosity and (sub)halo mass.  By
adjusting 3 free parameters, we attempt to simultaneously fit the
4 frequency bands of the {\it Planck} measurement of the CIB
anisotropy power spectrum.
To fit the data, we find that the star-formation efficiency must peak
on a halo mass scale of $\approx 5 \times 10^{12}~\msun$ and the
infrared luminosity per unit mass
must increase rapidly with redshift. By comparing our predictions
with a well-calibrated phenomenological model for shot noise, and with
a direct observation of source counts, we show that the mean duty
cycle of the underlying infrared sources must be near unity, indicating
that the CIB is dominated by long-lived quiescent star
formation, rather than intermittent short ``star bursts''.
Despite the improved flexibility of our model, the best simultaneous
fit to all four {\it Planck} channels remains relatively poor. We
discuss possible further extensions to alleviate the remaining tension
with the data. 
Our model presents a theoretical framework for a future
joint analysis of both background anisotropy and source count measurements.
\end{abstract} 
\begin{keywords} submillimetre: diffuse background -- submillimetre: galaxies -- galaxies: star formation -- galaxies: halos \end{keywords}

%%%%%%%%%%%%%%%%%%%%%%%%%%%%%%%%%%%%%%%%%%%%%%%%%%%%%%%%%%%%%%%%%%%%%%%%%%%%%%%%
\section{Introduction}
\label{sec:introduction}
%%%%%%%%%%%%%%%%%%%%%%%%%%%%%%%%%%%%%%%%%%%%%%%%%%%%%%%%%%%%%%%%%%%%%%%%%%%%%%%%

About half of the optical/UV emission from stars is absorbed by dust
in galaxies and re-emitted at infrared wavelengths
\citep{dwek98,fixsen98}. Measurements of the CIB therefore offer an
important channel to study the star-formation history of the
Universe. In particular, CIB measurements are sensitive to the star
formation out to high redshifts, due to the correlation between the
infrared luminosity of a galaxy and its star formation rate
\citep{kennicutt98}. Owing to the relatively poor angular resolution
of far-infrared telescopes, it is challenging to resolve and study
individual infrared sources.  Surveys to date are only able to resolve
the brightest sources, which are responsible for less than 15\% of the
total CIB \citep{oliver10b}.
Although the Atacama Large Millimeter/Submillimeter Array
(ALMA)\footnote{https://science.nrao.edu/facilities/alma} will soon be
able to resolve most 
of the background, it will only do so over a small region of the sky.
Large-scale fluctuations of the unresolved background will contain
complementary information that will remain useful even after the advent
of ALMA \citep{haiman00, knox01}.

Although fluctuations in the CIB were detected over a decade ago
\citep{LP00,Matsuhara+00}, these were limited to probing only the
Poisson contribution to the power spectrum.
Robust measurements of the clustered component in the CIB anisotropy,
probing the spatial correlations of the underlying sources, have
become possible relatively recently, and are now rapidly improving.
The first detection of clustered power came from 160$\,\mu$m {\it
Spitzer} data \citep{grossan07,lagache07}.  This was followed recently
by the Balloon-borne Large Aperture Submillimeter Telescope
(BLAST)\footnote{http://blastexperiment.info/}, which measured the
clustering power near the energetic peak of the CIB at 600, 860, and
1200$\,$GHz (500, 350, and $250\,\mu$m, respectively;
\citealt{viero09}).  \citet{hall10} reported the first detection of
clustered CIB power at millimeter wavelengths, in South Pole Telescope
(SPT)\footnote{http://pole.uchicago.edu/} data, soon followed by
\citet{dunkley10} in Atacama Cosmology Telescope
(ACT)\footnote{http://www.physics.princeton.edu/act} data, and
\citet{shirokoff11} with additional SPT data.  Most recently,
measurements of the CIB fluctuation power, extending over a broad
range in frequencies and angular scales, have been reported in
\cite{PlanckCIB} and \cite{amblard11}, inferred from {\it Planck} and
{\it Herschel} data, respectively.\footnote{See
http://www.rssd.esa.int/index.php?project=Planck and
http://herschel.esac.esa.int} In addition, a joint analysis of
multiple measurements has also been performed by \citet{addison11};
they find that the power spectra over the range of angular scales
and frequencies considered could be well fitted by a simple power law.

To make full use of this information, and to draw conclusions about
the cosmic star formation history, requires the development of a modeling
framework.  With too complex a model, degeneracies between model
parameters threaten to make meaningful conclusions impossible. On the
other hand, if the model is too simple, it may be unable to
accommodate all relevant information, such as observations of bright
individual sources -- or even worse: we may risk having conclusions
that are artificially driven by inadequate modeling assumptions.

The {\it Planck} and {\it Herschel} measurements, the most informative
to date, have so far been interpreted via halo models of large-scale
structure \citep{HaloModelReview}.  These models take advantage of the
relatively small uncertainty with which, given a model, statistical
properties of the spatial distribution of dark matter can be
calculated.  They do so by dividing the modeling into two stages.
First, the spatial distribution of dark matter is described (in
particular, by assuming that all dark matter is confined to lie with
spherically symmetric, collapsed halos). Second, the observable of
interest is related to the distribution of dark matter.  A key element
for the latter is the halo occupation distribution (HOD) function of
galaxies, which is a statistical description of how galaxies occupy
halos, depending on halo mass and redshift.

To avoid degeneracies that might arise in the presence of too many
free parameters, the halo model in \cite{PlanckCIB} and
\cite{amblard11}, to which we refer as the ``standard model'' in
  the following,
has been relatively simple -- ignoring many details 
of how infrared-luminous galaxies trace dark matter halos.  For
example, the HODs were assumed to be the same for all spectral types
and to not vary with redshift.  Even quite coarse details, such as the
dependence of the luminosity of a galaxy on halo mass have been
neglected.

These models have the virtue of simplicity, and without the data
demanding more complicated models, it would perhaps not be worth
abandoning such simplicity.  The primary motivation of our study,
however, is that the existing data are already difficult to understand
with these simple models.  For example, in the models used to
interpret the {\it Planck} CIB measurements, there is no single choice
of parameters that generates a good fit, simultaneously, to the power
spectra at all frequencies \citep{PlanckCIB}.  
Also, because these models do not include a dependence of the IR
luminosity on halo mass, neither the number counts of individual
sources (detectable at the bright end), nor the shot-noise power
levels can be calculated reliably.

In this paper, we extend the modeling framework used to date, by
allowing galaxies in halos with different masses to have different
luminosities.  We do so by assuming that satellite galaxies occupy
dark matter subhalos, and use a subhalo mass function derived from
cosmological N--body simulations.  The luminosity of each satellite
galaxy is then parametrically related to the mass of the subhalo.
Ours is not the first CIB model to allow for IR galaxies to have
different luminosities.  \citet{amblard07} discussed such a model,
based on the conditional luminosity function
\citep{cooray05,yang03,yang05}.  What is new in the present work is
the explicit association of IR galaxies with subhalos, and the use of
such a model for interpreting the {\it Planck} data. As we will show
below, the comparison of this model to the data indeed leads to novel
conclusions.

The modeling extension necessarily comes at the cost of a few
additional parameters, but has several benefits.  First, and perhaps
most importantly, it allows us to estimate a lower limit to the bright
end of the number counts.  Second, by introducing a duty cycle, and
comparing to source count observations, we can put limits on the duty
cycle.  In the context of our model, the data favor duty cycles near
unity. Physically, this implies that the bulk of the CIB is produced
by quiescent star formation, rather than by the intense, short
starbursts expected to be triggered in major mergers
\citep{Sanders+88,BH91}.  Third, there exists a tension between the
value of the power-law index $\alpha$, describing the dependence of
the number of satellite galaxies on halo mass, inferred either from
cosmological simulations, from optical galaxy data, or from the CIB
anisotropy measurements (see \S\ref{sec:illustration}).  Our model
naturally resolves this tension.

The remainder of this paper is organized as follows. In
\S\ref{sec:observations}, we review recent measurements of the CIB
anisotropy power spectrum by BLAST, SPT, and the space missions {\it
Herschel} and {\it Planck}.  In \S\ref{sec:model}, we first review
the basic formulae of the existing halo--based CIB models, and then
describe our extensions that include a more realistic $L-M$
relation. In \S\ref{sec:illustration}, we discuss the ingredients of
the extended model in more detail, and illustrate how each ingredient
affects the resulting CIB anisotropy power spectra. In
\S\ref{sec:mcmc}, we apply our model to the recent measurement by
{\it Planck}, and derive constrains on both the $L-M$ relation and on
the shape of the mean spectral energy distribution (SED) of the
sources. With enhanced flexibility, we attempt to simultaneously fit
the data at all measured frequencies.  In \S\ref{sec:shotnoise}, we
examine the shot noise levels and source counts computed with the new
model, compare them with an observationally-calibrated model and a
direct measurement, and place a constraint on the duty cycle of the
underlying infrared sources.  In \S\ref{sec:discussions}, we further
discuss the details of the fit, as well as several caveats and
potential future improvements to the model.  Finally, in
\S\ref{sec:conclusions}, we summarize our main conclusions.
Throughout this paper, we adopt the standard $\Lambda$CDM model as our
fiducial background cosmology, with parameter values consistent with
the seventh year {\it WMAP} results \citep{komatsu11}, i.e.,
\{$\Omega_m$,$\Omega_\Lambda$,$\Omega_b$,$h$,$\sigma_8$,$n_s$\} =
\{0.274, 0.726, 0.045, 0.705, 0.810, 0.96\}.

%%%%%%%%%%%%%%%%%%%%%%%%%%%%%%%%%%%%%%%%%%%%%%%%%%%%%%%%%%%%%%%%%%%%%%%%%%%%%%%%
\section{Measurements of CIB anisotropies}
\label{sec:observations}

Substantial progress has been made in measuring the power spectrum of
the CIB anisotropies in the past few years, thanks to the joint
efforts from BLAST, SPT, {\it Herschel} and {\it Planck}. 

In 2009, BLAST measured the power spectra in the GOODS-South field at
250, 350, and 500 $\mu m$ (i.e., 1200, 857, 600 GHz in frequencies,
respectively) in the multipole range of $940 \leq l \leq 10,800$
\citep{viero09}. After subtracting the Galactic cirrus and Poisson
noise, they found the variance of the CIB over the scales of 5' -- 25'
is consistent with a constant amplitude of $15\% \pm 1.7\%$ relative
to the CIB mean at all observing frequencies. The data could
alternatively be well fit by a linear halo bias model,
with bias parameters, $b=3.8\pm0.6, 3.9\pm 0.6$ and $4.4\pm 0.7$ at
1200, 857, 600 GHz, respectively. Interestingly, the data could not be
fit by the standard halo model based formula, with data points lying
below the model curves at small scales and above the model curves at
large scales. 
Later comparison with the {\it Planck} measurement shows a general
consistency between the two measurements. However, the data points
from the BLAST measurement are systematically higher at large scales.
This seems to suggest that the BLAST measurement have been contaminated
by residual Galactic cirrus at the large scales \citep{PlanckCIB}.
If true, the failure of the halo model fit, likely due to the
contamination of the data, does not necessarily mean the model is wrong.

\citet{hall10} reported SPT measurements of the auto- and
cross-correlation at 150 GHz and 220 GHz in the multipole range of
$2000 \leq l \leq 10000$. They also determined that the spectral
indices of the Poisson and clustered components between the two
observing frequencies are $3.86 \pm 0.23$ and $3.8 \pm 1.3$,
respectively.
This implies a steep slope of the SED of the dust emission (i.e. a
gray-body power-law index of $\beta \sim 2$; see definition in
eq.~\ref{eqn:thetanu} below).  Although the SPT measures the
small--scale clustering, a model based on linear halo bias
was still able to provide an acceptable fit to the data. This does not
necessarily mean the clustering is linear at such small scales, but
rather shows that the shape of non--linear power spectrum over the
scales probed by SPT can not be distinguished from that of the linear
power spectrum. A clear detection of the excess power from non--linear
clustering requires the measurement to cover the transition at
multiples of $l \sim 1000$. This is exactly the angular coverage
provided by {\it Herschel} and {\it Planck}.

Soon afterwards, \citet{dunkley10} published their
  measurement of the power spectra at 148 GHz
and 218 GHz, based on the ACT data collected during their 2008 season over
296 ${\rm deg^2}$. The clustered component of the IR power spectrum is
detected at $5 \sigma$, assuming an analytic model for its
shape. The spectral index of unresolved IR emission between the two
frequencies is found to be $3.69\pm 0.14$, consistent with the SPT measurement.

With the Spectral and Photometric Imaging Receiver (SPIRE) on board,
the {\it Herschel} Space Observatory, during its Science Demonstration
Phase, measured the power spectra of CIB anisotropies in the Lockman
Hole and GOODS South field at 250, 350, and 500 ${\mu m}$ over the
scales 1' -- 40' ($540 \lsim l \lsim 21 600$ in multipoles;
\citealt{amlard11}). The coverage of a wide range of angular scales,
combined with clean data, yielded a clear detection of the non--linear
clustering signal.  The standard halo model--based prescription
provides satisfactory fits to the data, but with different sets of
best-fit parameters at different frequencies. The minimum halo mass
for star--formation,
$M_{min}$, is constrained to be $\approx 3\times 10^{11}\msun$, which
the authors interpret as the most efficient mass--scale for star
formation. This mass--scale is lower than previous predictions by
semi-analytical models for galaxy formation \citep{gonzalez11}.

Most recently, the {\it Planck} team published their measurement of
the angular power spectra of CIB anisotropies, using maps of six
regions of low Galactic dust emission \citep[hereafter
A11]{PlanckCIB}. The power spectra were determined in 4 frequency
bands (217, 353, 545, and 857 GHz) over multipoles between $l=200$ and
$l=2000$. 
{\it Planck} found CIB anisotropies at the level $\Delta I/I=15\%$
compared to the mean CIB, and also that the fluctuating component and
the mean CIB have similar frequency spectra, consistent with the BLAST
results.
Unlike other missions, {\it Planck} was unable to independently
measure the amplitude of the shot noise, due to its relatively poor
angular resolution ($\sim 5$ arcmins). Instead, it adopted values
calculated using a parametric model by \citep[hereafter
B11]{bethermin11}. Again, different frequency bands yield
  different best-fit model parameters.

%%%%%%%%%%%%%%%%%%%%%%%%%%%%%%%%%%%%%%%%%%%%%%%%%%%%%%%%%%%%%%%%%%%%%%%%%%%%%%%%
\section{Halo model for CIB anisotropies and its extension}
\label{sec:model}

\subsection{Review of the existing models}
\label{subsec:model0}

The angular power spectrum of CIB anisotropies is defined through,
\begin{eqnarray}
<\delta I_{lm,\nu} \delta I^{}_{l^{\prime}m^{\prime},\nu^\prime}> =
C^{}_{\ell,\nu \nu^{\prime}} \delta_{ll^{\prime}}\delta_{mm^{\prime}},
\label{eqn:cll0}
\end{eqnarray}
where $\nu$ denotes the observing frequency and $I_{\nu}$ is the
measured specific intensity at that frequency. In a flat universe as
assumed throughout this paper, the specific intensity is related to
the comoving specific emission coefficient $j$ via
\begin{eqnarray}
\label{eqn:intensity}
I_{\nu}&=&\int dz\frac{d\chi}{dz} a j(\nu,z)\\\nonumber
&=&\int dz\frac{d\chi}{dz} a \bar{j}(\nu,z)\left(1+\frac{\delta
    j(\nu,z)}{\bar{j}(\nu,z)}\right),
\end{eqnarray}
where $\chi(z)$ is the comoving distance to redshift $z$, and
$a=1/(1+z)$ is the scale factor. Combining equations (\ref{eqn:cll0})
and (\ref{eqn:intensity}) and employing the Limber approximation
\citep{limber54}, we obtain
\begin{eqnarray}
C^{}_{\ell,\nu\nu^{\prime}}=\int\frac{dz}{\chi^2}\frac{d\chi}{dz} a^2
\bar{j}(\nu,z) \bar{j}(\nu^{\prime},z) P_{j,\nu\nu^{\prime}}^{}(k=l/\chi,z),
\label{eqn:cll1}
\end{eqnarray}
where $P_{j,\nu\nu^{\prime}}^{}$ is the 3-D power spectrum of the
emission coefficient, and is defined as follows,
\begin{eqnarray}
<\delta j(\vec{k},\nu) \delta j(\vec{k}^\prime,\nu^{\prime})> =
(2\pi)^3 \bar{j}(\nu)\bar{j}(\nu^{\prime}) P_{j,\nu\nu^{\prime}}^{}(\vec{k})\delta^3(\vec{k}-\vec{k}^{\prime}). 
\label{eqn:pj}
\end{eqnarray}
The existing models equate $P_j$ with the galaxy power spectrum
$P_{gal}$, assuming the CIB is sourced by galaxies, and that the
spatial variations in the emission coefficient trace the galaxy
number density,
\begin{eqnarray}
\delta j/\bar{j} = \delta n_{gal}/\bar{n}_{gal}.
\label{eqn:dj}
\end{eqnarray}
On large scales, in the linear regime, the galaxy power spectrum
follows the linear matter power spectrum, $P_{gal}=b^2 P_{lin}$, while
on small, non--linear scales, $P_{gal}$ is typically computed from the
halo model. In the framework of the halo model, the galaxy power
spectrum is a sum of three terms
\begin{eqnarray}
P_{gal}(k,z)=P_{1h}(k,z)+P_{2h}(k,z)+P_{shot}(k,z),
\label{eqn:pgal}
\end{eqnarray}
where $P_{1h}$ and $P_{2h}$ account for contributions from galaxy
pairs in the same halo and in different halos, respectively, and
$P_{shot}$ is the shot noise. The analytical expressions for $P_{1h}$
and $P_{2h}$ are \citep{HaloModelReview},
\begin{eqnarray}
\label{eqn:p1h}
P_{1h}(k,z)=\int dM
\frac{dN}{dM}(M,z)\times\\\nonumber
\frac{2N_{cen}(M)N_{sat}(M)u(k,z|M)+
N^2_{sat}(M) u^2(k,z|M)}{\bar{n}_{gal}^2}
\end{eqnarray}
and
\begin{eqnarray}
\label{eqn:p2h}
P_{2h}(k,z)=P_M(M,z)\times\\\nonumber
\left[\int dM \frac{dN}{dM}(M,z)\frac{N_{gal}(M,z) b(M,z)
u(k,z|M)}
{\bar{n}_{gal}}\right]^2,
\end{eqnarray}
where $M$ is the halo mass, $dN/dM$ is the halo mass function,
$u(k,z|M)$ is the Fourier transform of the halo density profile, and
$N_{cen}$ and $N_{sat}$, specified by the HOD, are the number of
central and satellite galaxies inside a halo, with
$N_{gal}=N_{cen}+N_{sat}$. Motivated by simulations
\citep[e.g.,][]{kravtsov04}, $N_{cen}$ is typically modeled as a
simple step function,
\begin{eqnarray}
  N_{cen} =
\left\{\begin{array}{ccc}
0& M<M_{min}\\
1&  M\ge M_{min} 
\end{array}\right.,
\label{eqn:ncen}
\end{eqnarray}
while $N_{sat}$ is parameterized by a power law,
\begin{eqnarray}
  N_{sat} = \left(\frac{M}{M_{sat}}\right)^\alpha.
\label{eqn:nsat}
\end{eqnarray}
Here, $M_{sat}$ is the arbitrary ``pivot'' halo mass that hosts one
satellite galaxy.  Note that in the above equations, $N_{sat}$,
$N_{cen}$, and $N_{gal}$ could be thought of as the {\em average}
number of galaxies inside halos with a fixed mass $M$, with negligible
scatter.  The parameterizations above have been extended to include
non--zero stochasticity in the relationship between halo mass and the
number of galaxies \citep[e.g.,][]{zheng05}.

\subsection{The improved model}
\label{subsec:model1}

The model above assumes that emissivity density traces galaxy number
density (equation \ref{eqn:dj}). This assumption implies that all
galaxies contribute equally to the emissivity density, {\it
irrespective} of the masses of their host halos. In other words, it
assumes that all galaxies have the same luminosity. 
Realistically,
however, both the luminosity and the clustering strength are closely
related to the mass of the host halo.  In particular, galaxies in
massive halos are likely both luminous and highly clustered. In
general, introducing a monotonic $L-M$ relation should give rise to
stronger clustering than the existing models, in which the more
strongly clustered sources (massive galaxies) have been assigned
artificially low luminosities. This improvement is indeed the main new
feature of our model.

In the following, we abandon the assumption of a mass--independent
luminosity, and use the standard halo model to directly compute the
power spectrum of the emission coefficient, $P_j$. Note that the
emission coefficient is related to the underlying galaxy population as
follows,
\begin{eqnarray}
j_{\nu}(z)&=&\int dL \frac{dn}{dL}(L,z)
\frac{L_{(1+z)\nu}}{4\pi}
\label{eqn:j0}
\end{eqnarray}
where $L$ denotes the infrared luminosity and $dn/dL$ is the infrared
galaxy luminosity function. Neglecting any scatter, the galaxy
luminosity is a function of the mass of the host dark matter halo (for
central galaxies), or subhalo (for satellite galaxies), and equation
(\ref{eqn:j0}) can be re-written as:
\begin{eqnarray}
\label{eqn:j1}
j_{\nu}(z)&=& \int dM \frac{dN}{dM}(z)\frac{1}{4\pi}\left[N_{cen}L_{cen,(1+z)\nu}(M,z)\right.\\
\nonumber 
& &\left.+\int dm \frac{dn}{dm}(M,z)L_{sat,(1+z)\nu}(m)\right],
\end{eqnarray}
where $m$ is the subhalo mass, and $dn/dm$ is the subhalo mass
function. Studies show that the luminosity of a satellite galaxy best
correlates with the mass or circular velocity of the host subhalo at
the time it is accreted into the main halo, i.e. before it looses
significant mass due to tidal stripping \citep{nagai05,vale06,
conroy06, wang06, wetzel10}. We therefore use this ``unstripped'' mass
to infer the infrared luminosity.  

The computation of the power spectrum of the emission coefficient
closely follows that of the galaxy power spectrum in the previous
section.  The only difference is that the galaxy numbers ($N_{cen}$
and $N_{sat}$) should be replaced by expressions accounting for
contributions from galaxies to the overall emissivity. Note that apart
from the additional weighting by luminosity, equation (\ref{eqn:j1})
is essentially identical to the expression for the number density:
\begin{eqnarray}
n(z)&=& \int dM \frac{dN}{dM}(z)[N_{cen}(M,z)+N_{sat}(M,z)].
\label{eqn:nz}
\end{eqnarray}
To compute $P_j$, we therefore only need to replace $N_{cen}$ by
\begin{eqnarray}
f_{\nu}^{cen}(M,z) = N_{cen}\frac{L_{(1+z)\nu}(M,z)}{4\pi},
\label{eqn:fcen}
\end{eqnarray}
and $N_{sat}$ by
\begin{eqnarray}
f_{\nu}^{sat}(M,z) = \int_{M_{min}}^{M}dm\frac{dn_{subhalo}}{dm}(m,z|M)\frac{L_{(1+z)\nu}(m,z)}{4\pi}
\label{eqn:fsat}
\end{eqnarray}
in equations (\ref{eqn:p1h}) and (\ref{eqn:p2h}). The final results are
\begin{eqnarray}
\label{eqn:pj1h}
P^{}_{1h,\nu\nu^{\prime}}(k,z)&=&\frac{1}{\bar{j}_{\nu}\bar{j}_{\nu^\prime}}\int_{M_{min}}^{\infty}dM\frac{dN}{dM}\\\nonumber
&&\times[f_{\nu}^{cen}(M,z)f_{\nu^\prime}^{sat}(M,z)u(k,M,z)
\\\nonumber
&&+f_{\nu^{\prime}}^{cen}(M,z)f_{\nu}^{sat}(M,z)u(k,M,z)\\\nonumber
&&+f_{\nu}^{sat}(M,z)f_{\nu^{\prime}}^{sat}(M,z)u^2(k,M,z)],\\
\label{eqn:pj2h}
P^{}_{2h,\nu\nu^{\prime}}(k,z)&=&\frac{1}{\bar{j}_{\nu}\bar{j}_{\nu^\prime}}D_{\nu}(k,z)D_{\nu^{\prime}}(k,z)P_{lin}(k,z),
\end{eqnarray}
where
\begin{eqnarray}
D_{\nu}(k,z)&=&\int_{M_{min}}^{\infty}dM\frac{dN}{dM}b(M,z)u(k,M,z))\\
\nonumber
&&\times [f_{\nu}^{cen}(M,z)+f_{\nu}^{sat}(M,z)].
\label{eqn:pjf}
\end{eqnarray}
It is straightforward to show that equations (\ref{eqn:pj1h}) and
(\ref{eqn:pj2h}) reduce to equations (\ref{eqn:p1h}) and
(\ref{eqn:p2h}) for a flat $L-M$ relation (galaxies of different
masses have the same luminosity), as assumed in previous models.

The main ingredients of this model, such as the halo mass function,
halo bias, and halo density profiles, have been carefully calibrated
using numerical simulations. Throughout this study, we define halos as
overdense regions with a mean density equal to 200 times the mean
density of the universe.
We assume an NFW profile \citep{navarro97} for the halo density, and
adopt the fitting function of \citet{tinker08} for the halo mass
function and its associated prescription for the halo bias
\citep{tinker10a}. For the subhalo mass function, we use the fitting
function of \citet[equation (12) in their paper]{tinker10b}.

We note that a luminosity--weighting scheme similar to the above have
been explored by \citet{sheth05} and \citet{skibba06}, and applied to
study resolved sources and the environment--dependence of galaxy
properties. To the best of our knowledge, this is the first time that
such a weighting scheme has been used for an analysis of unresolved
brightness fluctuations.

Finally, in addition to the clustering caused by correlated
large--scale structures, measurements of the CIB power spectrum
include shot noise from random fluctuations in the discrete number of
galaxies. In principle, this shot noise can be computed in the above
model, through the equation
\begin{eqnarray}
C_{l}^{shot}=\int_0^{S_{cut}}S^2\frac{dN}{dS}dS,
\label{eqn:cshot}
\end{eqnarray}
where $N$ includes both halos and subhalos, and $S$ is the source flux,
\begin{eqnarray}
S_{\nu}=\frac{a L_{(1+z)\nu}}{4 \pi \chi^2},
\label{eqn:flux}
\end{eqnarray}
and $S_{cut}$ is the flux above which individual sources are detected
and removed in a given experiment.  
Unfortunately, this computation
will likely remain inaccurate, given the simplicity of our model. In
particular, our model neglects any scatter in the $L-M$ relation, for
simplicity. This scatter is particularly relevant to the calculation
of the shot noise: as is clear from equation (\ref{eqn:cshot}), shot
noise involves the integration of $L^2$, which increases with scatter
for a fixed mean $L-M$ relation.  By neglecting scatter, our model
therefore underestimates the shot noise, and our result should be
understood as a lower limit. For this reason, we adopt the values
computed using the parametric model of B11 (their model successfully
fits bright source count measurements, and is also adopted by A11 as
the shot noise) when we fit the measured angular power spectra. The values
can be found in in Table (\ref{tbl:shotnoise}) below (and also in
Table~3 of A11). However, later in \S~\ref{sec:shotnoise}, equation
(\ref{eqn:cshot}) and (\ref{eqn:flux}) are employed to constrain the
duty cycle of the underlying sources.
%%%%%%%%%%%%%%%%%%%%%%%%%%%%%%%%%%%%%%%%%%%%%%%%%%%%%%%%%%%%%%%%%%%%%%%%%%%%%%%%
\section{Modeling the $L-M$ relation and its effect on CIB anisotropies}
\label{sec:illustration}

\subsection{Parameterizing the $L-M$ relation}
\label{subsec:L-M}

Before applying our model to the {\it Planck} data, we here specify
and discuss the model ingredients in detail. In the halo model, the
galaxy power spectrum is fully determined by the HOD, namely the
functions $N_{cen}(M,z)$ and $N_{sat}(M,z)$. In our model for the
background fluctuations, the power spectrum depends, additionally, on
the function $L_{(1+z)\nu}(m,z)$.  The latter depends on three
variables: the redshift $z$, the mass of the host (sub)halo, and the
observing frequency $\nu$. Limited by the current data quality, we
adopt a relatively simple form for $L_{(1+z)\nu}(m,z)$ in this study.
First, we do not make a distinction between halos and subhalos of the
same mass, i.e., we assume that the $L-M$ and $L-m$ relations are
identical.
Second, we assume, for simplicity, that all galaxies have the same
SED, independent of their masses and redshifts, and that the $L-M$
relation does not evolve with redshift except for an overall
normalization.  The dependence of $L$ on the three variables is then
separable, and $L_{(1+z)\nu}(m,z)$ can be written as the product
\begin{eqnarray}
L_{(1+z)\nu}(m,z)=L_0 \Phi(z) \Sigma(m) \Theta[(1+z)\nu],
\label{eqn:lfunc}
\end{eqnarray}
where $L_0$ is a normalization factor.  In the following, we discuss
each of the three components in more detail.

\vspace{\baselineskip}
\noindent {\bf (1) Redshift evolution $\Phi(z)$} 

For a given (sub)halo mass, the luminosity and the star formation rate
(SFR) are expected to increase with redshift, because of the higher
gas accretion rates, higher gas fractions, and more compact geometries
at early times. Additionally, mergers, which can trigger starbursts,
are more frequent at high redshift.  In this study, we parameterize
$\Phi(z)$ as a power law,
\begin{eqnarray}
\Phi (z)= \left(1+z\right)^{s_z}.
\label{eqn:phiz}
\end{eqnarray}
This is partly motivated by the study of the specific star formation
rate (sSFR; SFR per unit stellar mass). If we assume the stellar mass
to halo mass ratio evolves only mildly with redshift (supported by
semi-analytical studies such as \citealt{neistein11}), the sSFR should
have a redshift evolution similar to that of $L_\mathrm{IR}/M$ due to
a correlation between SFR and infrared luminosity \citep{kennicutt98}.

The evidence for a smooth power--law evolution of the sSFR with
redshift is somewhat mixed. Semi-analytical models of
galaxy formation indeed show that the redshift evolution of sSFR,
assumed to be primarily driven by the fresh gas supply, could
be well fit by a power--law with a slope of $\sim 2.5$
\citep{neistein08, dekel09, oliver10a}. Observations, however, suggest a
more complicated shape, with a steep evolution below $z\lsim 2$, and a
plateau beyond this redshift (see \citealt{bouche10} and
\citealt{weinmann11} for a compilation of measurements). Such a
transition is hard to explain from a theoretical perspective
\citep{bouche10, weinmann11}. Due to the different shapes, theoretical
models tend to predict a sSFR lower by a factor of a few at $z\sim 2$
and much higher at $z>4$ than inferred from observations, even though
these models could more or less reproduce the redshift evolution of
the overall SFR \citep{weinmann11}.
On the other hand, there are large uncertainties in current
observations, especially at high redshifts. For example, measurements
by \citet{yabe09} and \citet{schaerer10} suggest the sSFR continue to
rise beyond $z>2$. Given these uncertainties, we consider two
scenarios: $s_z$ has a single value at all redshifts (case 0), and
$s_z$ is fixed to 0 at $z>2$ (case 1).

\begin{figure*}
\begin{tabular}{c}
\rotatebox{-90}{\resizebox{80mm}{!}{\includegraphics{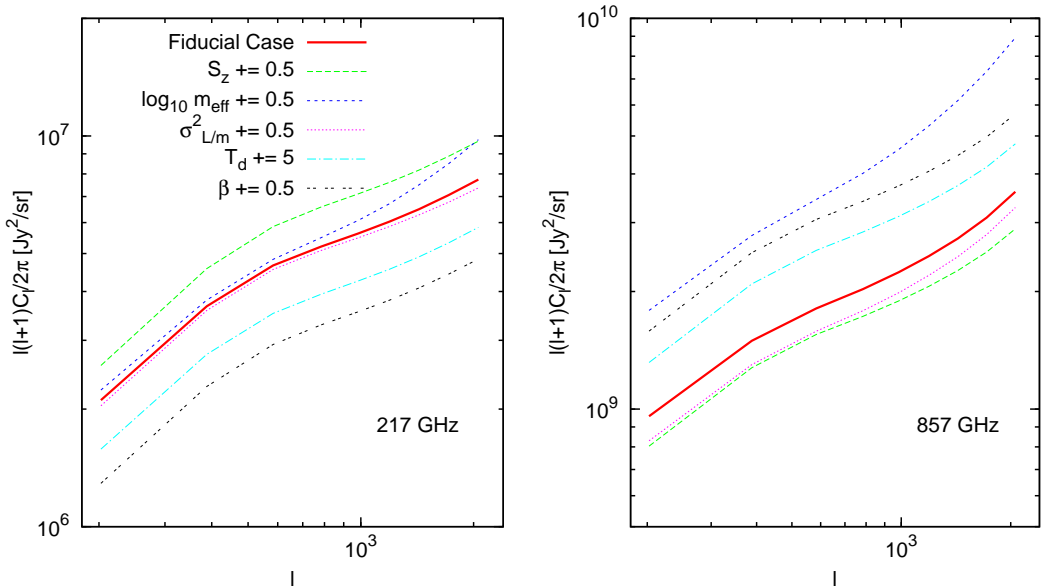}}}
\end{tabular}
\caption{CIB power spectra predicted by our models at 217 GHz (left
  panel) and 857 (right panel). The solid curve in both panels
  corresponds to our fiducial model, with parameter values of
  $s_{z}=3$, $m_{eff}=10^{12.5}~\msun$, $\sigma^2_{L/m}=0.5$,
  $T_{d}=34$ K, and $\beta=2$ (see \S\ref{subsec:L-M} for the
  definitions of these parameters). The other curves show how the
  power spectrum is modified when an individual parameter is changed
  by the amount indicated in the labels.}
\label{fig:ill}
\end{figure*}

\vspace{\baselineskip}
\noindent {\bf (2) L-M relation $\Sigma(m)$}

A robust conclusion from comparisons between observed galaxy
luminosity functions and halo mass functions is that star formation is
only effective in a certain range of halo masses.  In particular,
star-formation must be suppressed at both low and high halo masses --
for example, by feedback processes such as photoionization heating,
supernovae, active galactic nuclei (AGN) and virial shocks
\citep{birnboim03, keres05, dekel06, bower06, croton06}.  The quantity
of particular interest is the halo mass scale at which the star
formation is most efficient, corresponding to a peak in the
luminosity-to-mass ($L/m$) ratio. Assuming, for simplicity, that $L/m$
as a function of mass $m$ is log--normal, we model the $L-m$ relation
as follows,
\begin{eqnarray}
\Sigma (m)= m \frac{1}{\sqrt{2\pi \sigma_{L/m}^2}}
\exp\left[{-\frac{(\log_{10}m-\log_{10}m_{eff})^2}{2\sigma_{L/m}^2}}\right].
\label{eqn:sigmam}
\end{eqnarray}
Here $m_{eff}$ describes the peak of the specific IR emissivity per
unit mass, and $\sigma_{L/m}$ describes the range of halo masses for
producing IR luminosity.  At the low--mass end, in addition to the
exponential drop in equation (\ref{eqn:sigmam}), we impose a minimum
mass of $10^{11}\msun$, i.e., $L=0$ for $m<10^{11}\msun$. This cutoff
is motivated by the work of \citet{bouche10}. Later, we will see the
best-fit $m_{eff}$ is more than 2$\sigma_{L/m}$ away from the cutoff,
so effects of applying such a cutoff are mild.

\vspace{\baselineskip}
\noindent {\bf (3) SED shape $\Theta(\nu)$}

We adopt the same form for the SED as \citet{hall10}. The SED is a
graybody at low frequencies, and is a power law at high frequencies,
\begin{eqnarray}
\Theta (\nu) \propto
\left\{\begin{array}{ccc}
\nu^{\beta}B_{\nu}(T_d)& \nu<\nu_0\\
\nu^{-\gamma}&  \nu\ge \nu_0 
\end{array}\right.,
\label{eqn:thetanu}
\end{eqnarray}
where $B_{\nu}$ is the Planck function and $T_d$ is an effective dust
temperature. These two functions are connected smoothly at the
frequency $\nu_0$ that satisfies
\begin{eqnarray}
\frac{d {\rm ln}[\nu^{\beta}B_{\nu}(T_d)]}{d{\rm ln}\nu}=-\gamma.
\end{eqnarray}
We fix $\gamma=2$ as in \citet{hall10}, but, unless stated otherwise,
we  allow $\beta$ and $T_d$ to be free parameters.

\subsection{Effects on CIB anisotropies}
\label{subsec:impact}

Next, we illustrate how the CIB anisotropy power spectrum depends on
each of the model parameters. For this purpose, we adopt the following
set of fiducial model parameter values: $s_{z}=3$,
$m_{eff}=10^{12.5}~\msun$, $\sigma^2_{L/m}=0.5$, $T_{d}=34$ K,
$\beta=2$.
The choice of $s_{z}=3$ is based on low redshift measurements of the
sSFR \citep{noeske07, dunne09, oliver10a, rodighiero10}; while
$m_{eff}=10^{12.5}~\msun$ is around the peak of the stellar-mass to halo
mass ratio found by the abundance matching technique \citep[e.g.,][]{guo10}. 
The value $T_{d}=34$ K is consistent with
measurements such as \citet[$T_d=35.6\pm 4.9$ K]{dunne00},
\citet[$T_d=36 \pm 7$ K]{chapman05}, and \citet[$T_d=28\pm 8$
K]{amblard10}. The assumption $\beta = 2$ follows the common practice
in the literature, and is supported by many theoretical
considerations \citep[e.g.,][]{draine84, mathis89}. The
choice $\sigma^2_{L/m}=0.5$ implies that the
``efficient'' mass range for star--formation covers about one and half
orders of magnitude in halo mass. Though this number is somewhat
arbitrary, it is unlikely to severely bias the final results, since,
as demonstrated below, the power spectra are relatively insensitive to
$\sigma^2_{L/m}$. 

We next vary one parameter at a time, and examine how the power
spectrum changes compared to the fiducial case. The overall
normalization $L_0$ is fixed using the amplitude of the mean CIB
measured by {\it FIRAS} (this assumption will be discussed and relaxed
in \S\ref{sec:discussions} below).  More specifically, we choose
$L_0$ such that the ratio of the predicted and measured mean
intensity, averaged over all frequency bands, is equal to unity,
\begin{eqnarray}
\frac{1}{N_{f}}\sum_{i=1}^{N_{f}} \frac{I^{}_{FIRAS,\nu_i}}{I^{}_{model,\nu_i}}=1.
\label{eqn:norm}
\end{eqnarray}
Here $N_{f}$ is the number of frequency bands in which the power
spectra have been measured (for the most recent {\it Planck}
measurement by A11, $N_{f} =4$), $I^{}_{model,\nu_i}$ is the intensity
at frequency $\nu_i$ predicted by our model while $I^{}_{FIRAS,\nu_i}$
is the same quantity measured by the {\it FIRAS} instrument. For the
latter, we use the values quoted in \citet{gispert00}.

The results are shown in Figure~\ref{fig:ill} for the highest (857
GHz) and lowest (217 GHz) frequency bands of {\it Planck}. The power
spectra in the fiducial model are plotted as solid thick curves,
against which the other curves should be compared.  There are two
distinct ways in which the parameters can affect the power spectra:
either by changing the shape of the spectrum ($C^{}_{\ell,\nu\nu}
\propto I_\nu^2$) or by changing the effective bias factor ($b$).
If the parameter enters primarily through the spectrum, then it must
have opposite effects at high and low frequencies, since the average
intensity is kept fixed by the normalization
(equation~\ref{eqn:norm}).  On the other hand, if the parameter enters
primarily through the bias factor, then it has similar effects at
different frequencies.  This distinction helps us recognize how the
parameters affect the power spectra, and also indicate their
degeneracies -- parameters which enter the same way are more likely to
be degenerate.  Figure~\ref{fig:ill} reveals that three of the
parameters, $s_z$, $T_d$ and $\beta$, affect the spectrum, while the
other two parameters, $m_{eff}$ and $\sigma^2_{L/m}$, enter through
the bias factor. Consequently, degeneracies are likely strong among
$s_z$, $T_d$ and $\beta$, and between $m_{eff}$ and $\sigma^2_{L/m}$.

Interestingly, increasing $m_{eff}$ not only changes the overall
clustering strength, but also increases the amplitude of the
non-linear (1-halo term) contribution, relative to the linear (2-halo
term) contribution to the power spectrum. This is seen clearly as the
upturn in both panels of Figure~\ref{fig:ill} in the short--dashed
[blue] curve at $\ell \gsim 10^3$, and is similar to the effect of
increasing the value of the parameter $\alpha$ in the previous models
(see equation \ref{eqn:nsat}).  In this sense, $m_{eff}$ differs from
$M_{min}$, since the latter
mostly affects the overall clustering. This feature of our model
arises from the non-flatness of the $L-M$ relation, and is worth
discussing further, because, as we will see below, the data requires
significant power on small scales, which can be provided by the 1-halo
term.

For the intensity fluctuations, what matters for the relative strength
of the 1-halo and 2-halo terms is the total luminosity, rather than
total number, of satellite galaxies. To be more concrete, let us
suppose that the total luminosity of satellite galaxies inside a halo
of mass $M$ can be parameterized by a power--law, similar to
equation~(\ref{eqn:nsat}),
\begin{eqnarray}
  L_{tot,sat} = \left(\frac{M}{M_{sat}}\right)^{\alpha_L}.
\label{eqn:lsat}
\end{eqnarray}
It is then ${\alpha_L}$, rather than $\alpha$, that determines the
small scale clustering.  If the luminosities of galaxies were
independent of the masses of their host (sub)halos, $\alpha_L$ would
be equal to $\alpha$. However, for a more realistic $L-M$ relation,
for which luminosity and mass are positively correlated over the
relevant mass range, $\alpha_L$ is larger than $\alpha$.  This is
because in addition to the number of subhalos increasing with the halo
mass, the average sub-halo mass increases, as well. Consequently,
satellite galaxies are, on average, brighter in more massive halos,
rendering $\alpha_L$ larger than $\alpha$.

\begin{figure}
\begin{tabular}{c}
\rotatebox{-90}{\resizebox{60mm}{!}{\includegraphics{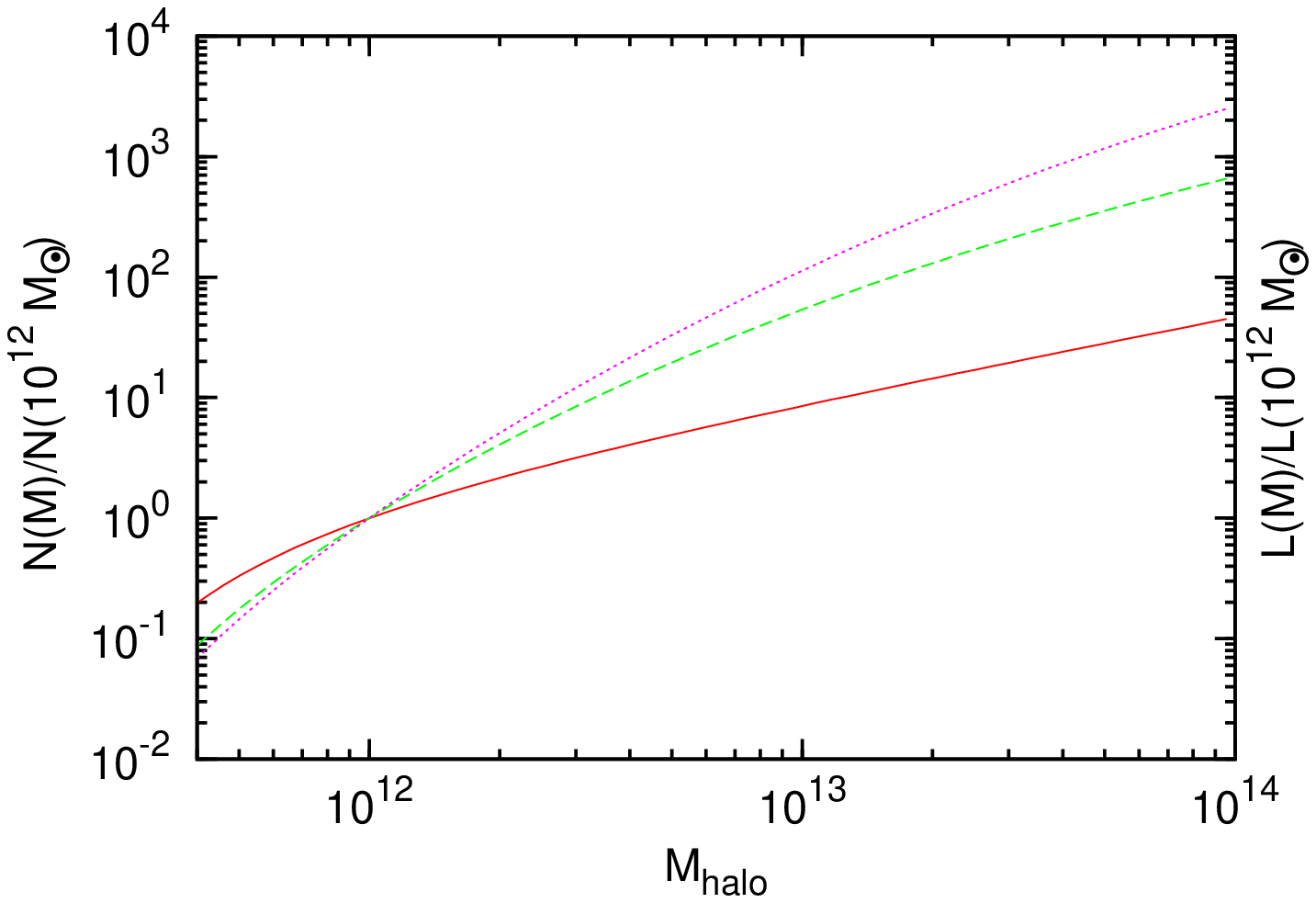}}}
\end{tabular}
\caption{The solid curve shows the number of subhalos more massive
  than $10^{11}~\msun$ inside a halo of mass $M_{halo}$, while the
  dashed and dotted curves show the total luminosity of the these
  subhalos. All curves are normalized by their values at
  $M_{halo}=10^{12}~\msun$, and $m_{eff}$ is taken to be
  $10^{12.5}~\msun$ and $10^{13}~\msun$ for the dashed and dotted
  curves, respectively.  The figure illustrates that the total
  luminosities of the satellites increase more rapidly than their
  number; this feature of our model helps to explain the large CIB
  power observed on small angular scales.}
\label{fig:alphal}
\end{figure}

To illustrate this point explicitly, in Figure~\ref{fig:alphal} we
show the total number (solid curve) and the total luminosity (dashed
and dotted curves) of satellite galaxies inside a halo of mass
$M_{halo}$. All curves are shown normalized by their values at
$M_{halo}=10^{12}~\msun$. As expected, the total luminosity increases
more rapidly with halo mass than the number of satellites. Further
comparing the dashed ($m_{eff}=10^{12.5}~\msun$) and the dotted
($m_{eff}=10^{13}~\msun$) curves shows that the slope of the relation
itself increases with $m_{eff}$.  This explains why the amplitude of
the 1-halo term increases with $m_{eff}$ compared to the 2-halo term.

In their recent analysis of the CIB power spectrum, \citet{amblard11}
have found that in order to fit the (large) observed power on small
angular scales, the number of satellite galaxies had to be
increased compared to that expected from numerical simulations and
optical surveys.  In particular, the expectation is $\alpha \lsim 1$
and $M_{sat}=10-25 M_{min}$ \citep[e.g.,][]{gao04, kravtsov04,zheng05,
hansen09},
while the analysis of \citet{amblard11} requires either $M_{sat}=3.3
M_{min}$ or, effectively, $\alpha >1$ (reducing $M_{sat}$ and
increasing $\alpha$ have similar effects on the power spectrum since
both increase the number of satellites and therefore raise the small
scale power).  Our model naturally resolves this tension by
distinguishing $\alpha$ from $\alpha_L$, and, as we will demonstrate
below, is able to fit the data without the presence of any additional
low--mass satellite galaxies.

%%%%%%%%%%%%%%%%%%%%%%%%%%%%%%%%%%%%%%%%%%%%%%%%%%%%%%%%%%%%%%%%%%%%%%%%%%%%%%%%
\section{Constraints from {\it Planck}}
\label{sec:mcmc}
%%%%%%%%%%%%%%%%%%%%%%%%%%%%%%%%%%%%%%%%%%%%%%%%%%%%%%%%%%%%%%%%%%%%%%%%%%%%%%%%

\begin{table}
  \caption{\em Best-fit values and marginalized 1-$\sigma$ errors, as
    well as the reduced $\chi^2$ from fits to the {\it Planck} CIB
    power spectra. Within the parenthesis are the numbers of degrees
    of freedom. Five variants of our model are considered.  Case
    0: simultaneous fit to the 4 frequency bands of {\it Planck}, with
    three of the five model parameters allowed to vary, as explained
    in \S\ref{sec:model}. Case 1: same as case 0, except $s_z=0$
    enforced at $z>2$. Case 2: same as case 0, except the 857 GHz
    channels is ignored.  Case 3: the graybody SED slope $\beta$ is
    fixed at 2, and the dust temperature $T_d$ is instead a free
    parameter. Case 4: the normalization $L_0$ is allowed to vary (and
    found to be constrained to the range $L_0=(0.73\pm
    0.05)L_{\star}$, where $L_{\star}$ is the solution of
    eq.~\ref{eqn:norm}).}
    \label{tbl:fit}
\begin{center}
\begin{tabular}{l c c c c}
\hline \hline
 & $s_z$ & $log_{10} m_{eff}$ & $\beta$ & reduced $\chi^2$\\
\hline
Case 0 &$   4.64^{+0.13}_{-0.10}$  &$12.65^{+0.14}_{-0.15}$  &$
2.68^{+0.09}_{-0.08}$ &1.82 (33)\\ 
Case 1 &$  5.57^{+0.99}_{-0.96}$ &$ 12.65^{+0.09}_{-0.08}$ & $
1.90^{+0.07}_{-0.07}$&2.78 (33)\\ 
Case 2 &$  0.93^{+0.65}_{-0.65}$ &$ 12.78^{+0.20}_{-0.22}$ &$
1.71^{+0.15}_{-0.14}$ &1.12 (24)\\ 
\hline
\hline
 & $s_z$ & $log_{10} m_{eff}$ & $T_d$ & reduced $\chi^2$\\
\hline
Case 3 &$  4.85^{+0.80}_{-0.79}$ &$ 12.73^{+0.23}_{-0.23}$ &$
44.81^{+7.09}_{-6.78}$ &2.43 (33)\\ 
Case 4 &$  6.48^{+0.52}_{-0.44}$ &$ 13.56^{+0.17}_{-0.17}$ &$
43.49^{+3.43}_{-3.32}$ & 1.70 (32)\\
\hline
\end{tabular}
\end{center}
\end{table}

\begin{figure*}
\begin{tabular}{c}
\rotatebox{-90}{\resizebox{100mm}{!}{\includegraphics{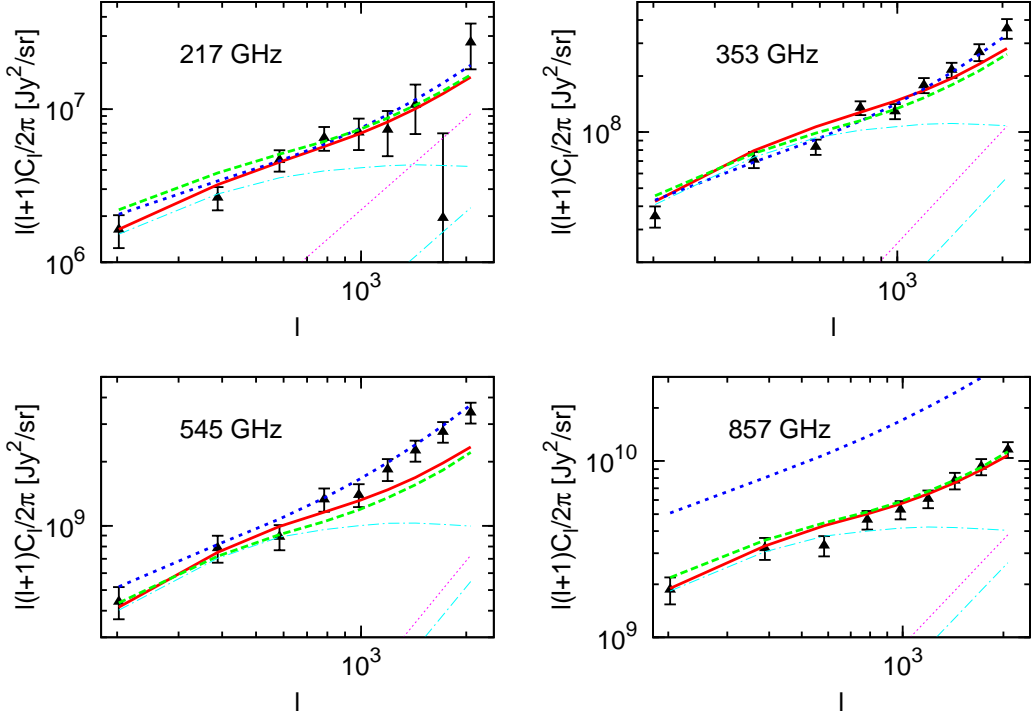}}}
\end{tabular}
\caption{CIB power spectra at 217 GHz, 353 GHz, 545 GHz, and
  857 GHz measured by {\it Planck} (data points with error bars; taken from A11)
  and predicted in our best--fit models (curves). The solid [red],
  long-dashed [green] and short-dashed [blue] curves show the total
  power spectra, including the shot noise, in case 0, 1 and 2,
  respectively (see Table 1 and \S\ref{sec:mcmc} for a summary of the
  models). For reference, the contributions from the shot noise
  (dotted curves), the 1-halo term (straight dash-dotted lines at
  $\ell > 10^3$) and the 2-halo term (dash-dotted curves) are shown
  separately in the fiducial ``case 0''.  }
\label{fig:bestfit}
\end{figure*}

We are now ready to derive constraints from the {\it Planck}
data. Given the flexibility of our extended model, we attempt to fit
the CIB power spectra in the four {\it Planck} frequency bands
simultaneously. As mentioned above, however, there are strong
degeneracies among our 5 model parameters, which prevent us from
simultaneously constraining each.  We therefore first proceed by
fixing two of the parameters -- the dust temperature $T_d$, and the
mass-range for significant luminosity, $\sigma^2_{L/m}$ -- at their
fiducial values. In addition, we also apply a flat prior on $s_z$, $0<
s_z < 7$.  Hereafter, this will be referred to as ``case 0''; we will
explore other possibilities below.

To obtain the best-fit values and their confidence levels, we adapted
the {\it CosmoMC} Monte Carlo Markov Chain (MCMC) code
\citep{lewis02}, and applied it to the data in the four {\it Planck}
channels presented in A11.  The chains are checked visually and
diagnosed using the statistical tests provided by {\it CosmoMC} to
ensure their convergence.
The best-fit values and the marginalized errors are listed in the
first row of Table~\ref{tbl:fit} and the power spectrum in the
best-fit model is shown explicitly at each of the four frequencies by
the solid [red] curves in Figure~\ref{fig:bestfit}.  As this table and
the figure demonstrates, with our admittedly crude assumptions, and
varying only three parameters, we are able to obtain
  good fits to 
the {\it Planck} data, with the exception of the high $\ell$ data points
at 545 GHz.  where we underpredict the observed power.

The best-fit $s_z\approx 4.6$ is qualitatively consistent with the
rapid evolution of the sSFR indicated by recent measurements. As we
will see below, the uncertainties induced by our model assumptions are
likely much larger than the statistical errors. It is nevertheless
interesting that the data, without any direct measurements of
redshifts, requires a fairly rapid evolution of the sSFR.

The halo mass scale for most efficient star formation is constrained
to be around $m_{eff}\approx 10^{12.65}~\msun$. This is consistent
with the typical masses of the host halos of submillimetre galaxies,
as predicted by semi-analytical models
\citep[e.g.][]{gonzalez11}. This mass scale is also in general
agreement with that inferred from the clustering measurements of
resolved bright sources \citep[e.g.,][]{cooray10}.  These studies all
converge and indicate that star--formation is most vigorous in halos
with masses of a few $\times 10^{12} ~\msun$.  It is therefore
interesting to note that a recent measurement of the small-scale CIB
power by {\it Herschel} derived a much lower mass--scale of $3\times
10^{11}~\msun$.  \cite{amlard11}
There are two possible reasons for this difference. First, the
measurements themselves might be inconsistent. A recent comparison
between the two measurements indeed shows that the power spectra
measured by {\it {\it Planck} } at 545 GHz and 857 GHz are higher than
those measured by {\it Herschel}. Second, as mentioned in the
Introduction, the meaning of $M_{min}$ in the old model (adopted by
the {\it Herschel} team) does not correspond to that of the $m_{eff}$
defined here, invalidating a direct comparison between the two.

The best-fit value for $\beta$ ($\approx 2.7$) is larger than the value commonly
assumed ($\beta=2$).  The common assumption is due to the fact that
simple models for both insulating and conducting materials naturally give $\beta = 2$ at long wavelengths \citep{draine84, gordon88}.  However, long-range disorder in the dust grains can lead to
$\beta > 2$ \citep{meny07}.  \citet{meny07} also refers to observational and laboratory evidence 
for a wide range of $\beta$ values, including those greater than 2.  Note that others have also inferred a steep spectral index for the background anisotropy spectrum at long wavelengths \citep{hall10, dunkley10, shirokoff11}.

Inferring $\beta$ from observations is complicated by contributions from dust grains at varying temperatures.  A mixture of cold and hot dust flattens the spectrum in the wavelength range between the two spectral peaks; if analyzed with a single-temperature grey-body model, the inferred value of $\beta$ will thus be artificially low as emphasized by \citet[e.g.,][]{reach95}.

Just as a model that is missing a {\em cold} component can lead to an
artificially low inference of $\beta$, missing a hot component can
lead to an artificially {\em high} inference of $\beta$.  What we
infer as a high value of $\beta$, may be due to a component, not
included in our model, with a high effective temperature where $T_{\rm
  eff} \equiv T_D/(1+z)$.  Perhaps we are missing a significant low-redshift
component.

More importantly, we
emphasize that our best--fit $\beta$ value depends especially
sensitively on model assumptions.  In particular, as discussed above,
$\beta$ is degenerate with $s_z$; we will shortly see that $\beta$ is
reduced to $\sim 2$ if we make a different assumptions about the
redshift-evolution.  Another degeneracy of $\beta$ is with $T_d$; a
possible cause of a large $\beta$ is that we adopted the wrong dust
temperature.  The value $T_d = 34$K is motivated by the temperature
measurements in resolved sources, which, on average, are brighter, and
reside at lower redshift, than the fainter sources responsible for the
unresolved background.  Finally, the measurement errors are
considerable ($\gsim 5$K).

For the reasons in the preceding paragraph, we next consider the case
where we fix $\beta=2$, and allow the dust temperature to vary
instead. The results are listed in Table~\ref{tbl:fit} under ``case
3''.  The values of $s_z$ and $log_{10} m_{eff}$ are consistent with
those from case 0, while the dust temperature is found to be around 45
K, $\approx$11 K higher than assumed in case 0. This numerical result
confirms a rough estimate based on Figure~\ref{fig:ill}: a reduction
of 0.68 in $\beta$ could be compensated by an increase of $\sim 10$ K
in $T_d$.  The best-fit dust temperature is then somewhat higher than
the measurements quoted above (although within their 2$\sigma$
errors).  One could indeed speculate that at the higher redshifts
probed by the CIB measurement, galaxies are more compact, and dust in
their interstellar medium is hotter, since dust particles reside
closer to the stars.

\begin{figure}
\begin{tabular}{c}
\rotatebox{-0}{\resizebox{80mm}{!}{\includegraphics{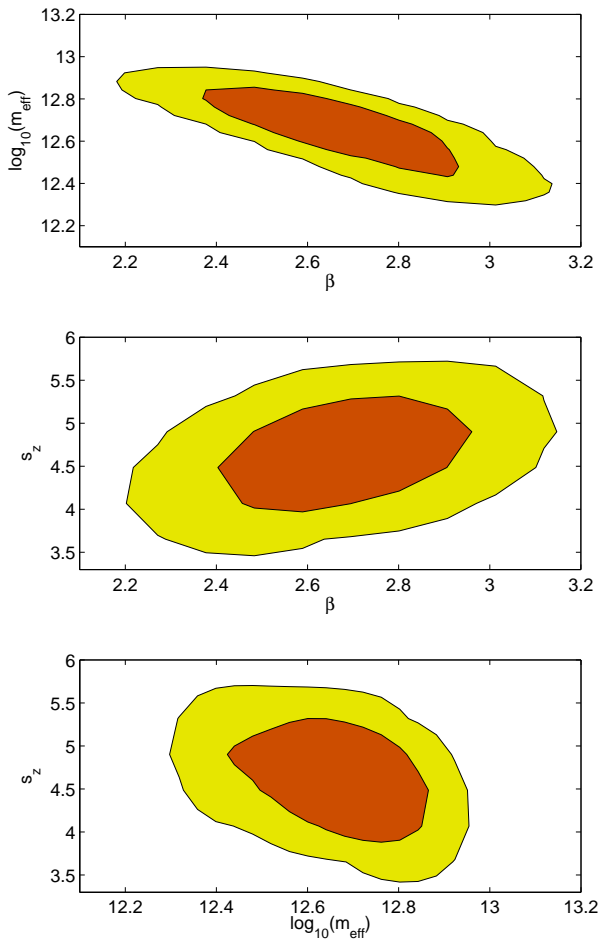}}}
\end{tabular}
\caption{Constraints on the parameters $s_z$, $m_{eff}$ and $\beta$
  derived from Monte Carlo Markov Chains, by simultaneously fitting
  the CIB power spectrum in the four {\it Planck} frequency bands (in
  the fiducial model ``case 0'' defined in \S\ref{sec:mcmc}). The
  inner [red] and outer [yellow] contours show the 68\% and 95\%
  confidence levels, respectively.  In each panel, the third parameter
  has been marginalized over.}
\label{fig:contour}
\end{figure}

In Figure~\ref{fig:contour}, we show 2-D confidence contours of the
parameters in case 0.  These the correlations between the 3 free
parameters. In particular, $m_{eff}$ is anti-correlated with $s_z$ and
$\beta$, while $s_z$ and $\beta$ are positively correlated with each
other. These correlations can be understood by examining the
dependence of the power spectrum on individual parameters, shown in
Figure~\ref{fig:ill} above.

Finally, owing to the current uncertainty on whether the sSFR
continues to rise beyond redshift $z>2$ (as discussed above), we
consider a variation on our fiducial model with a high-$z$ plateau.
In this variant (referred to as ``case 1''), we allow the $L-M$
normalization to increase with redshift as before, i.e. as a
power--law with a constant $s_z$ (eq.\ref{eqn:phiz}) at low
redshift, but we set $s_z=0$ at $z>2$.
The results for this case are shown in the second row of
Table~\ref{tbl:fit}, and by the dashed curves in
Figure~\ref{fig:bestfit}. The best-fit $s_z$ and $\beta$ both change
considerably, with $s_z$ increasing to $s_z\approx 5.6$, and
$\beta\approx 1.9$ reduced close to its expected value of
$\beta\approx 2$. This indicates that uncertainties in $s_z$ and
$\beta$ due to model assumptions are large.  Interestingly, the
best-fit $m_\mathrm{eff}$ remains almost unchanged. The minimum
$\chi^2$ in case 1 is somewhat larger than in case 0 (2.78 v.s. 1.82),
indicating that the data favor case 0.

%%%%%%%%%%%%%%%%%%%%%%%%%%%%%%%%%%%%%%%%%%%%%%%%%%%%%%%%%%%%%%%%%%%%%%%%%%%%%%%%
\section{Shot noise and the duty-cycle of the underlying infrared source}
\label{sec:shotnoise}
%%%%%%%%%%%%%%%%%%%%%%%%%%%%%%%%%%%%%%%%%%%%%%%%%%%%%%%%%%%%%%%%%%%%%%%%%%%%%%%%

As mentioned in \S\ref{subsec:model1} above, our model can be used to
obtain a lower limit on the shot noise.  As we show here, this enables
us to place a lower limit on the the duty-cycle of the underlying
infrared sources, $f_{duty}$, and to thus speculate on their nature
(intermittent or quiescent).  For simplicity, throughout the
discussion below, we will assume a universal, constant duty cycle,
independent of redshift and luminosity.

Combining equations~(\ref{eqn:cshot}) and (\ref{eqn:flux}) implies
that
\begin{eqnarray}
C^{shot}\propto L_{act}^2 N_{act},
\label{eqn:cshotill}
\end{eqnarray}
where $L_{act}$ is the luminosity of ``active'' galaxies, and
$N_{act}$ is the number of active galaxies. For a fixed abundance of
galaxies, $N_{act}$ is proportional to the duty-cycle, and if the
total luminosity density ($L_{act} N_{act}$) is fixed, then $L_{act}$
is inversely proportional to it. Thus, the amplitude of the shot noise
power is inversely proportional to the duty-cycle, $C^{shot}\propto
1/f_{duty}$.\footnote{This neglects the fact that the brightest
sources have been masked. For {\it Planck}, the flux cut is large (a
few hundred mJy), and we find that the inverse relation between
$C^{shot}$ and $f_{duty}$ holds accurately as long as $f_{duty}>0.1$.}

In Table~\ref{tbl:shotnoise}, we list the shot noise estimates in our
models, in Case 0 and Case 1 (the conclusions from the other cases are
very similar) and compare these with results from the empirical model of
B11. The values of $S_{cut}$ needed in the calculation can be found
in Table 3 of A11.
The B11 shot noise estimates have been carefully calibrated with
source count measurements. The model employs a double exponential
function to parametrize the luminosity function, and double power-laws
to parameterize its redshift evolution. By adjusting 13 free
parameters, the model successfully reproduces the counts from the
mid-infrared to the millimeter wavelengths, as well as the
mid-infrared luminosity functions. 
The duty cycle is taken to be either $f_{duty}=1$ (mimicking the value
for quiescent star--formation in long--lived galaxies) or
$f_{duty}=0.01$ (mimicking short--lived star--bursts, as expected when
star--formation is triggered by major mergers).  As the table shows,
the shot noise levels for $f_{duty}=1$ are of the same order as the
results of B11 (although our shot noise levels for Case 0(1) are
somewhat higher(lower) than those of B11).  On the other hand, with
$f_{duty}=0.01$, the shot noise in our models severely exceed those of
B11 (by factors of $\gsim20$). Given this vast difference, we conclude
that {\it the average duty cycle of the underlying sources is on the
order of unity}. This suggests that the infrared background is mainly
contributed by normal quiescent galaxies, rather than starbursts. This
conclusion is consistent with recent numerical simulations
\citep[e.g.,][]{hopkins10}, which find that $\sim 85\%$ of the
infrared background is contributed by normal galaxies.

It is interesting to note that between the two cases, our Case 1 has a
lower shot noise, due to its flatter redshift evolution. With the
total intensity fixed, the shot noise decreases with increasing source
number density. In Case 1, a higher fraction of the CIB originates
from low redshift, where the galaxy number density is higher.

It might appear worrisome that our lower limits on the shot noise, in
Case 0, are slightly higher than those of B11.  This, however, could
arise because our assumed $\sigma_{L/m}^2$ is too small.  Increasing
$\sigma_{L/m}^2$ would reduce the average luminosity, and consequently
reduce the shot noise levels.  If this were the case, then $m_{eff}$
in Table~\ref{tbl:fit} would have been underestimated. On the other
hand, comparisons with the BLAST measurement show that the shot noise
estimates by the model of B11 could be lower than the true values by a
considerable amount (60\% at 545 GHz, 20\% at 857 GHz; see A11).

\begin{table}
  \caption{Shot noise levels obtained by B11, compared with the lower
    limits computed in our models, for Cases 0 and 1, and for two
    different values of the duty--cycle $f_{duty}$.}
    \label{tbl:shotnoise}
\begin{center}
\begin{tabular}{l c c c c c c}
\hline \hline
&$f_{duty}$ & 217 GHz & 353 GHz & 545 GHz & 857 GHz\\
\hline
%B11& - & $13.8 \pm 2.9$&  $159 \pm 22$ & $1078 \pm 92$ & $ 5646 \pm
%367$ \\\hline
B11& - & $12.2 \pm 2.9$&  $138 \pm 22$ & $1150 \pm 92$ & $ 5923 \pm
367$ \\\hline
\multirow{2}{*}{Case 0}& 1 &  14.79  & 308.39  & 2305.94  & 7874.02  \\
&0.01& 516.93  & 5790.21  & 34813.30  & 85269.09 \\\hline
%& 136.51  & 2779.14  & 18755.38  & 73557.02\\\hline
\multirow{2}{*}{Case 1}&1 &   5.71  & 114.43  & 1057.59  & 5434.20 \\
&0.01& 570.56  & 7884.97  & 72353.10  & 89765.42  \\\hline
%&0.1&  57.06  & 1144.30  & 10576.25  & 54343.75\\\hline
%\multirow{2}{*}{Case 2}&1 &   9.19  & 207.37  & 2785.62  & 32652.63\\ 
%&0.01& 659.11  & 12752.90  & 28567.77  & 73975.74\\\hline
%&0.1&  91.86  & 2073.68  & 27855.90  & 123125.63\\
\end{tabular}
\end{center}
\end{table}

\begin{figure}
\begin{tabular}{c}
\rotatebox{-90}{\resizebox{60mm}{!}{\includegraphics{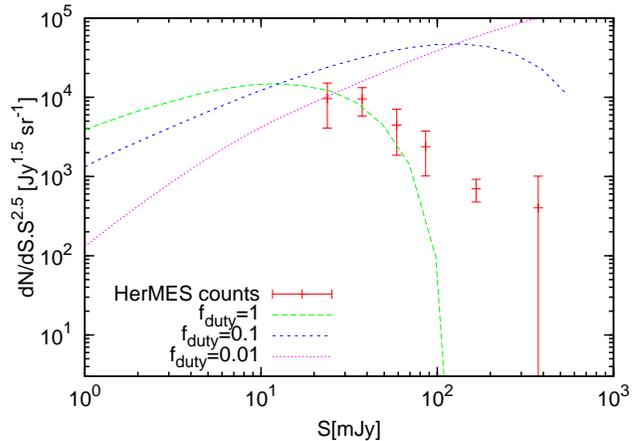}}}
\end{tabular}
\caption{Resolved bright source counts at 857 GHz from the HerMES
  survey (data points) are compared with lower limits derived from our
  models, assuming $f_{duty}=1$ (long dashed [green] curve), 0.1
  (short-dashed [blue] curve) and 0.01 (dotted [red] curve). The other
  model parameters are set to their best-fit values in Case 0.}
\label{fig:counts}
\end{figure}

In principle, our IR source population model could be compared
directly to measurements of the counts of resolved bright sources, as
well.  The distribution of source counts as a function of flux,
especially at the bright end, also depends sensitively on the assumed
the duty cycle. In particular, if the total CIB intensity is kept
fixed, then the number of bright sources increases, while the number
of faint sources decreases with decreasing duty-cycle.\footnote{This
can be easily verified by visualizing shifting a log$N$-log$S$ diagram
``downward'' and then toward higher luminosities.}  Because we do not
include any scatter in the $L-M$ relation, which would increase the
number of bright sources, the source counts predicted by our model are
lower limits at the bright end.  By comparing the model predictions
with measurements, we are therefore able to derive a lower limit on
the duty-cycle. We show the results of such a comparison at 857 GHz in
Figure~\ref{fig:counts}. The data points are taken from the {\it
Herschel} Multi-tiered Extra-galactic (HerMES) survey
\citep{oliver10b}, and the long-dashed, short-dashed, and dotted
curves are the lower limits on the counts, computed in our model for
$f_{duty}=1$, 0.1 and 0.01, respectively (with the other parameters
fixed at their best--fit values in Case 0).  This figure clearly shows
that only $f_{duty}=1$ is consistent with the observed number counts;
the other two values significantly overpredict the number of the
bright sources (by more than an order of magnitude).  We therefore
arrive at the same conclusion as that drawn from the analysis of the
shot noise, namely that the CIB must be produced primarily by
long-lived sources with $f_{duty}\approx 1$.

%%%%%%%%%%%%%%%%%%%%%%%%%%%%%%%%%%%%%%%%%%%%%%%%%%%%%%%%%%%%%%%%%%%%%%%%%%%%%%%%
\section{Discussion}
\label{sec:discussions}
%%%%%%%%%%%%%%%%%%%%%%%%%%%%%%%%%%%%%%%%%%%%%%%%%%%%%%%%%%%%%%%%%%%%%%%%%%%%%%%%

In this section, we discuss a few remaining issues and caveats related
to our main results.

\vspace{\baselineskip}
\noindent {\bf (1) Small scale clustering at 545 GHz}

Although we are able to fit all four {\it Planck} channels 
 with an overall reduced $\chi^2$ of 1.82, formally this is still
a bad fit, and our best-fit model
noticeably underpredicts the small-scale power at 545 GHz (by a factor
of $\sim$two).  There are at least two possible ways to alleviate this
problem.

First, the shot noise, which is almost parallel to the 1-halo term
over the scales considered in this study, is uncertain. The model
could match the data better if the shot noise were raised by a
suitable amount. Indeed, as we mentioned previously, the shot noise
levels measured by BLAST \citep{viero09} are about $60\%$ and $20\%$
higher than the model predictions by B11 at 545 GHz and 857 GHz,
respectively (A11). If we keep the best--fit model parameters in Case
0 as listed in the first row of Table~\ref{tbl:fit}, but increase the
shot noise at 545 GHz and 857 GHz by these factors, the tension
between the model and data at the small scales is significantly
 reduced.

Second, one can certainly attempt to modify and/or fine-tune our
best--fit models, to increase the small--scale power.  As we mentioned
in \S\ref{sec:illustration}, this could be achieved by increasing
$m_{eff}$. However, we have found no simple way to increase the small
scale clustering at 545 GHz, without affecting other frequencies,
especially at 857 GHz.  In particular, this is because there is
considerable overlap between different frequencies, caused by the
extended range of redshifts over which sources contribute to the CIB.
To illustrate this inter--dependence of frequencies, we perform a fit
to the three lowest frequency bands, ignoring the 857GHz channel
(denoted as ``Case 2'' in the third row of Table~\ref{tbl:fit}, and
shown with short dashed curves in Figure~\ref{fig:bestfit}).  The
best fit to these three channels yields a $\chi^2$ of 27 for 24 degrees of
freedom.  The power at 857 GHz in this case is, however, overpredicted by
a factor of $\approx$two.  We conclude that non-trivial modifications
to the SED, or other ingredients of our model would be needed in order
to improve this situation.

\vspace{\baselineskip}
\noindent {\bf (2) The normalization of the $L-M$ relation}

Thus far, the normalization of the $L-M$ has been fixed by
equation~(\ref{eqn:norm}). This makes the MCMC chains converge more
easily, and yields tighter constraints on other parameters. However,
fixing $L_0$ with infinite precision could raise two possible
concerns.  First, while equation (\ref{eqn:norm}) guarantees that our
model is {\it on average} consistent with the existing {\it FIRAS}
measurement of the mean CIB, individual frequencies might still
deviate, with a few frequencies far below the measured values, while
the others far above. Second, since the uncertainties of the {\it
FIRAS} measurement are non-negligible, could our conclusions be
altered if the normalization is allowed to vary?

\begin{figure}
\begin{tabular}{c}
\rotatebox{-90}{\resizebox{60mm}{!}{\includegraphics{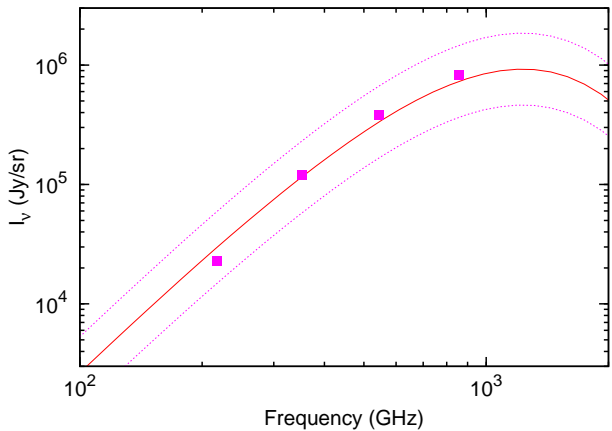}}}
\end{tabular}
\caption{The spectrum of the mean CIB in our fiducial Case 0, shown at
    the four frequency bands of {\it Planck} (solid squares).  The
    spectra in our other models is similar. Our mean background is
    consistent with the background spectrum measured by {\it FIRAS}
    (solid curve), lying well within the $\pm1\sigma$ allowed region
    (indicated by the dotted curves; \citealt{gispert00}).}
\label{fig:intensities}
\end{figure}

To address the first concern, in Figure~\ref{fig:intensities} we show
the spectrum of the mean CIB, as predicted in our model (for Case 0
only, as the spectra in the other cases are very similar).  This
figure also shows the {\it FIRAS} measurement, along with the
$\pm1\sigma$ region (as derived from the errors on the parameters in
the {\it FIRAS} fitting formula; \citealt{gispert00}).
Clearly, the intensities at all frequencies are consistent with the
observations.

\begin{figure}
\begin{tabular}{c}
\rotatebox{-0}{\resizebox{80mm}{!}{\includegraphics{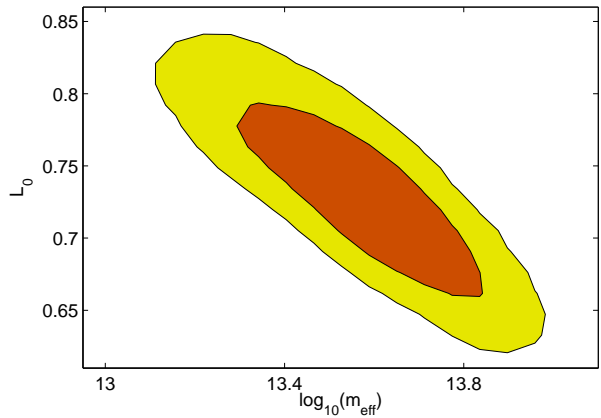}}}
\end{tabular}
\caption{Confidence contours (68\% and 95\%) on $log_{10} m_{eff}$ and
  $L_0$ for Case 4, in which the normalization $L_0$ was allowed to
  vary.  The contours have been marginalized over the other two free
  parameters, $s_z$ and $T_d$.}
\label{fig:contour2}
\end{figure}

To address the second concern, we relaxed the normalization $L_0$ in
the Monte-Carlo fitting. Other settings were kept the same as in Case
3, except that we imposed the prior $m_{eff}\leq 10^{14}~\msun$, in
addition to the prior on $s_z$.  This was necessary because otherwise
the fits allowed combinations of unphysically high $m_{eff}$ with
suitably low normalizations.
With the addition of $L_0$, we then have a total of four free
parameters. Only the {\it Planck} data was used in the fitting,
although in principle, the {\it FIRAS} data could be used explicitly
for the normalization, as well. The marginalized errors on the
parameters are shown in Table~\ref{tbl:fit}, marked ``Case 4''.  The
best--fit value for the normalization $L_0$ is found to be $(0.73 \pm
0.05)L_{\star}$, where $L_\star$ is the normalization satisfying
equation~(\ref{eqn:norm}). Compared to other cases, $m_{eff} \sim
10^{13.6} \, \msun$ is considerably higher and well beyond the
  halo mass of a typical galaxy. The large $m_{eff}$ raises the small scale clustering,
and helps alleviate the small scale tension we discussed previously. It
also increases the large--scale clustering (by a smaller factor),
which is then compensated by the lower normalization $L_0$. The
negative correlation between $m_{eff}$ and $L_0$ is also clear from
the confidence contours shown in Figure~\ref{fig:contour2}.
Interestingly, the best--fit normalization, $\sim 0.73~{I_{FIRAS}}$,
is within the allowed range of the {\it FIRAS} measurement, despite
the fact that this was not imposed on the fits.

\vspace{\baselineskip}
\noindent {\bf (3) Future improvements of the model}

Although successfully describing the {\it Planck} data, and an
improvement over previous version, our model is still admittedly very
simplified. In the following, we give a few examples how the model
could be improved and extended to include more physical ingredients.

1) As mentioned above, we neglect any scatter in the $L-M$ relation.
This scatter could be modeled, and would impact the estimates of the
shot noise and the duty--cycle, in particular. The scatter itself
might be mass-- and redshift--dependent.

2) We have a single galaxy population in our model. In reality,
multiple populations, such as normal blue galaxies, star--bursts and
obscured AGNs may all make appreciable contributions to the CIB. These
populations might have different SEDs, duty cycles, $L-M$ relations
and redshift evolutions. The fraction of starbursts and AGNs could be
taken from theoretical merger--tree models, calibrated by existing
data on the IR galaxy populations. In addition, the various properties
of each population may depend not only on redshift, but also on
environment.

3) In the current model, very simple forms have been assumed for the
SED and $L-M$ relations, and the SED was not allowed to evolve with redshift.
These are features that could be improved in future work. For example,
an (asymmetric) double power--law with a suitable set of parameters is
probably a better representative of the true $L-M$ relation, though it
requires more parameters.

4) Central and satellite galaxies could have been treated with
different $L-M$ relations.

With such improvements, the model could be developed so as to properly
compute the shot noise, and simultaneously fit number counts of
resolved sources and the power spectra of background fluctuations.

%%%%%%%%%%%%%%%%%%%%%%%%%%%%%%%%%%%%%%%%%%%%%%%%%%%%%%%%%%%%%%%%%%%%%%%%%%%%%%%%
\section{Conclusions}
\label{sec:conclusions}
%%%%%%%%%%%%%%%%%%%%%%%%%%%%%%%%%%%%%%%%%%%%%%%%%%%%%%%%%%%%%%%%%%%%%%%%%%%%%%%%

In this study, we developed a halo--model based formalism to compute
the power spectrum of the cosmic infrared background (CIB). Although
previous, similar, models provided excellent fits to individual 
frequency bands, they were based on an unrealistic assumption that the
luminosities of galaxies are independent of the host halo 
masses.
We relaxed this assumption by incorporating the subhalo mass
function, together with a more realistic $L-M$ relation, into our
model.  With these improvements, we were able to naturally resolve the
tension between the high power in the CIB on small angular scales
(requiring $\alpha>1$, where the number of satellite galaxies $N_{\rm
sat}$ in a halo scales as $\propto M^{\alpha}$) and optical data and
numerical simulations results (which favor $\alpha\lsim 1$).

With our improved model, we were also able to simultaneously fit all
four frequency bands of the {\it Planck} measurement of the CIB
anisotropy spectrum, by varying only three parameters: $s_z$ (the
describing the redshift evolution of the $L-M$ relation), $m_{eff}$
(the halo mass--scale on which star--formation is most efficient), and
$\beta$ (a parameter describing the graybody SED of the unresolved IR
sources).  We found that star formation is most efficient in halos
with relatively high masses of $m_{eff}=$several$\times
10^{12}~\msun$, in general agreement with semi-analytical models of
galaxy formation. We also found that the {\it Planck} data favor an
increase in the $L-M$ normalization with redshift.  Finally, by
comparing the shot noise and bright source counts predicted by our
models with an observationally calibrated empirical model and a direct
measurement, respectively, we conclude that the duty-cycle of the
source is on the order of unity.  This implies that the CIB is
dominated by long--lived, quiescent star-forming galaxies, rather than
short--lived star--bursts. This conclusion should be robust, since our
estimates of shot noise and bright source counts are already
conservative, owing to our neglect of any scatter in the $L-M$
relation.

The model we presented here, with the potential to compute the shot
noise and number counts, in addition to the correlated background
fluctuations, provides a theoretical framework for a future joint
analysis of the unresolved CIB background and resolved source
counts. Such a coherent analysis will no doubt afford new insight into
the relation between dust--enshrouded star--formation and dark matter
halo properties.

\vspace{-0.5\baselineskip}

\section*{Acknowledgments}

We thank Bruce Draine, Nicholas Hall, Olivier Dore, Guilanine Lagache, Carlo
Giocoli and Dan Marrone for useful discussions. We acknowledge the use of the CNSI
Computer Facilities at UC Santa Barbara for majority of the numerical
work. LK and SPO acknowledge NSF grants AST 0709498 and AST 0908480
respectively.

%\vspace{-2\baselineskip}

\bibliography{cib}

\begin{thebibliography}{73}
\expandafter\ifx\csname natexlab\endcsname\relax\def\natexlab#1{#1}\fi

\bibitem[{{Addison} {et~al}\mbox{.}(2011){Addison}, {Dunkley}, {Hajian},
  {Viero}, {Bond}, {Das}, {Devlin}, {Halpern}, {Hincks}, {Hlozek}, {Marriage},
  {Moodley}, {Page}, {Reese}, {Scott}, {Spergel}, {Staggs}, \&
  {Wollack}}]{addison11}
{Addison} G.~E. {et~al.}, 2011, ArXiv e-prints

\bibitem[{{Ade} {et~al}\mbox{.}(2011){Ade}, {Aghanim}, {Arnaud}, {Ashdown},
  {Aumont}, {Baccigalupi}, {Balbi}, {Banday}, {Barreiro}, \&
  et~al.}]{PlanckCIB}
{Ade} P.~A.~R. {et~al.}, 2011, ArXiv e-prints (A11)

\bibitem[{{Amblard} \& {Cooray}(2007)}]{amblard07}
{Amblard} A., {Cooray} A., 2007, \apj, 670, 903

\bibitem[{{Amblard} {et~al}\mbox{.}(2011{\natexlab{a}}){Amblard}, {Cooray},
  {Serra}, {Altieri}, {Arumugam}, {Aussel}, {Blain}, {Bock}, {Boselli}, {Buat},
  {Castro-Rodr{\'{\i}}guez}, {Cava}, {Chanial}, {Chapin}, {Clements}, {Conley},
  {Conversi}, {Dowell}, {Dwek}, {Eales}, {Elbaz}, {Farrah}, {Franceschini},
  {Gear}, {Glenn}, {Griffin}, {Halpern}, {Hatziminaoglou}, {Ibar}, {Isaak},
  {Ivison}, {Khostovan}, {Lagache}, {Levenson}, {Lu}, {Madden}, {Maffei},
  {Mainetti}, {Marchetti}, {Marsden}, {Mitchell-Wynne}, {Nguyen}, {O'Halloran},
  {Oliver}, {Omont}, {Page}, {Panuzzo}, {Papageorgiou}, {Pearson},
  {P{\'e}rez-Fournon}, {Pohlen}, {Rangwala}, {Roseboom}, {Rowan-Robinson},
  {Portal}, {Schulz}, {Scott}, {Seymour}, {Shupe}, {Smith}, {Stevens},
  {Symeonidis}, {Trichas}, {Tugwell}, {Vaccari}, {Valiante}, {Valtchanov},
  {Vieira}, {Vigroux}, {Wang}, {Ward}, {Wright}, {Xu}, \& {Zemcov}}]{amblard11}
{Amblard} A. {et~al.}, 2011{\natexlab{a}}, \nat, 470, 510

\bibitem[{{Amblard} {et~al}\mbox{.}(2011{\natexlab{b}}){Amblard}, {Cooray},
  {Serra}, {Altieri}, {Arumugam}, {Aussel}, {Blain}, {Bock}, {Boselli}, {Buat},
  {Castro-Rodr{\'{\i}}guez}, {Cava}, {Chanial}, {Chapin}, {Clements}, {Conley},
  {Conversi}, {Dowell}, {Dwek}, {Eales}, {Elbaz}, {Farrah}, {Franceschini},
  {Gear}, {Glenn}, {Griffin}, {Halpern}, {Hatziminaoglou}, {Ibar}, {Isaak},
  {Ivison}, {Khostovan}, {Lagache}, {Levenson}, {Lu}, {Madden}, {Maffei},
  {Mainetti}, {Marchetti}, {Marsden}, {Mitchell-Wynne}, {Nguyen}, {O'Halloran},
  {Oliver}, {Omont}, {Page}, {Panuzzo}, {Papageorgiou}, {Pearson},
  {P{\'e}rez-Fournon}, {Pohlen}, {Rangwala}, {Roseboom}, {Rowan-Robinson},
  {Portal}, {Schulz}, {Scott}, {Seymour}, {Shupe}, {Smith}, {Stevens},
  {Symeonidis}, {Trichas}, {Tugwell}, {Vaccari}, {Valiante}, {Valtchanov},
  {Vieira}, {Vigroux}, {Wang}, {Ward}, {Wright}, {Xu}, \& {Zemcov}}]{amlard11}
---, 2011{\natexlab{b}}, \nat, 470, 510

\bibitem[{{Amblard} {et~al}\mbox{.}(2010){Amblard}, {Cooray}, {Serra}, {Temi},
  {Barton}, {Negrello}, {Auld}, {Baes}, {Baldry}, {Bamford}, {Blain}, {Bock},
  {Bonfield}, {Burgarella}, {Buttiglione}, {Cameron}, {Cava}, {Clements},
  {Croom}, {Dariush}, {de Zotti}, {Driver}, {Dunlop}, {Dunne}, {Dye}, {Eales},
  {Frayer}, {Fritz}, {Gardner}, {Gonzalez-Nuevo}, {Herranz}, {Hill}, {Hopkins},
  {Hughes}, {Ibar}, {Ivison}, {Jarvis}, {Jones}, {Kelvin}, {Lagache}, {Leeuw},
  {Liske}, {Lopez-Caniego}, {Loveday}, {Maddox}, {Micha{\l}owski}, {Norberg},
  {Parkinson}, {Peacock}, {Pearson}, {Pascale}, {Pohlen}, {Popescu},
  {Prescott}, {Robotham}, {Rigby}, {Rodighiero}, {Samui}, {Sansom}, {Scott},
  {Serjeant}, {Sharp}, {Sibthorpe}, {Smith}, {Thompson}, {Tuffs}, {Valtchanov},
  {van Kampen}, {van der Werf}, {Verma}, {Vieira}, \& {Vlahakis}}]{amblard10}
---, 2010, \aap, 518, L9+

\bibitem[{{Barnes} \& {Hernquist}(1991)}]{BH91}
{Barnes} J.~E., {Hernquist} L.~E., 1991, \apjl, 370, L65

\bibitem[{{B{\'e}thermin} {et~al}\mbox{.}(2011){B{\'e}thermin}, {Dole},
  {Lagache}, {Le Borgne}, \& {Penin}}]{bethermin11}
{B{\'e}thermin} M., {Dole} H., {Lagache} G., {Le Borgne} D., {Penin} A., 2011,
  \aap, 529, A4+ (B11)

\bibitem[{{Birnboim} \& {Dekel}(2003)}]{birnboim03}
{Birnboim} Y., {Dekel} A., 2003, \mnras, 345, 349

\bibitem[{{Bouch{\'e}} {et~al}\mbox{.}(2010){Bouch{\'e}}, {Dekel}, {Genzel},
  {Genel}, {Cresci}, {F{\"o}rster Schreiber}, {Shapiro}, {Davies}, \&
  {Tacconi}}]{bouche10}
{Bouch{\'e}} N. {et~al.}, 2010, \apj, 718, 1001

\bibitem[{{Bower} {et~al}\mbox{.}(2006){Bower}, {Benson}, {Malbon}, {Helly},
  {Frenk}, {Baugh}, {Cole}, \& {Lacey}}]{bower06}
{Bower} R.~G., {Benson} A.~J., {Malbon} R., {Helly} J.~C., {Frenk} C.~S.,
  {Baugh} C.~M., {Cole} S., {Lacey} C.~G., 2006, \mnras, 370, 645

\bibitem[{{Chapman} {et~al}\mbox{.}(2005){Chapman}, {Blain}, {Smail}, \&
  {Ivison}}]{chapman05}
{Chapman} S.~C., {Blain} A.~W., {Smail} I., {Ivison} R.~J., 2005, \apj, 622,
  772

\bibitem[{{Conroy}, {Wechsler} \& {Kravtsov}(2006){Conroy}, {Wechsler}, \&
  {Kravtsov}}]{conroy06}
{Conroy} C., {Wechsler} R.~H., {Kravtsov} A.~V., 2006, \apj, 647, 201

\bibitem[{{Cooray} {et~al}\mbox{.}(2010){Cooray}, {Amblard}, {Wang},
  {Arumugam}, {Auld}, {Aussel}, {Babbedge}, {Blain}, {Bock}, {Boselli}, {Buat},
  {Burgarella}, {Castro-Rodriguez}, {Cava}, {Chanial}, {Clements}, {Conley},
  {Conversi}, {Dowell}, {Dwek}, {Eales}, {Elbaz}, {Farrah}, {Fox},
  {Franceschini}, {Gear}, {Glenn}, {Griffin}, {Halpern}, {Hatziminaoglou},
  {Ibar}, {Isaak}, {Ivison}, {Khostovan}, {Lagache}, {Levenson}, {Lu},
  {Madden}, {Maffei}, {Mainetti}, {Marchetti}, {Marsden}, {Mitchell-Wynne},
  {Mortier}, {Nguyen}, {O'Halloran}, {Oliver}, {Omont}, {Page}, {Panuzzo},
  {Papageorgiou}, {Pearson}, {Perez Fournon}, {Pohlen}, {Rawlings}, {Raymond},
  {Rigopoulou}, {Rizzo}, {Roseboom}, {Rowan-Robinson}, {Schulz}, {Scott},
  {Serra}, {Seymour}, {Shupe}, {Smith}, {Stevens}, {Symeonidis}, {Trichas},
  {Tugwell}, {Vaccari}, {Valtchanov}, {Vieira}, {Vigroux}, {Ward}, {Wright},
  {Xu}, \& {Zemcov}}]{cooray10}
{Cooray} A. {et~al.}, 2010, \aap, 518, L22+

\bibitem[{{Cooray} \& {Milosavljevi{\'c}}(2005)}]{cooray05}
{Cooray} A., {Milosavljevi{\'c}} M., 2005, \apjl, 627, L89

\bibitem[{{Cooray} \& {Sheth}(2002)}]{HaloModelReview}
{Cooray} A., {Sheth} R., 2002, \physrep, 372, 1

\bibitem[{{Croton} {et~al}\mbox{.}(2006){Croton}, {Springel}, {White}, {De
  Lucia}, {Frenk}, {Gao}, {Jenkins}, {Kauffmann}, {Navarro}, \&
  {Yoshida}}]{croton06}
{Croton} D.~J. {et~al.}, 2006, \mnras, 365, 11

\bibitem[{{Dekel} \& {Birnboim}(2006)}]{dekel06}
{Dekel} A., {Birnboim} Y., 2006, \mnras, 368, 2

\bibitem[{{Dekel}, {Sari} \& {Ceverino}(2009){Dekel}, {Sari}, \&
  {Ceverino}}]{dekel09}
{Dekel} A., {Sari} R., {Ceverino} D., 2009, \apj, 703, 785

\bibitem[{{Draine} \& {Lee}(1984)}]{draine84}
{Draine} B.~T., {Lee} H.~M., 1984, \apj, 285, 89

\bibitem[{{Dunkley} {et~al}\mbox{.}(2010){Dunkley}, {Hlozek}, {Sievers},
  {Acquaviva}, {Ade}, {Aguirre}, {Amiri}, {Appel}, {Barrientos}, {Battistelli},
  {Bond}, {Brown}, {Burger}, {Chervenak}, {Das}, {Devlin}, {Dicker}, {Bertrand
  Doriese}, {Dunner}, {Essinger-Hileman}, {Fisher}, {Fowler}, {Hajian},
  {Halpern}, {Hasselfield}, {Hernandez-Monteagudo}, {Hilton}, {Hilton},
  {Hincks}, {Huffenberger}, {Hughes}, {Hughes}, {Infante}, {Irwin}, {Juin},
  {Kaul}, {Klein}, {Kosowsky}, {Lau}, {Limon}, {Lin}, {Lupton}, {Marriage},
  {Marsden}, {Mauskopf}, {Menanteau}, {Moodley}, {Moseley}, {Netterfield},
  {Niemack}, {Nolta}, {Page}, {Parker}, {Partridge}, {Reid}, {Sehgal},
  {Sherwin}, {Spergel}, {Staggs}, {Swetz}, {Switzer}, {Thornton}, {Trac},
  {Tucker}, {Warne}, {Wollack}, \& {Zhao}}]{dunkley10}
{Dunkley} J. {et~al.}, 2010, ArXiv e-prints

\bibitem[{{Dunne} {et~al}\mbox{.}(2000){Dunne}, {Eales}, {Edmunds}, {Ivison},
  {Alexander}, \& {Clements}}]{dunne00}
{Dunne} L., {Eales} S., {Edmunds} M., {Ivison} R., {Alexander} P., {Clements}
  D.~L., 2000, \mnras, 315, 115

\bibitem[{{Dunne} {et~al}\mbox{.}(2009){Dunne}, {Ivison}, {Maddox},
  {Cirasuolo}, {Mortier}, {Foucaud}, {Ibar}, {Almaini}, {Simpson}, \&
  {McLure}}]{dunne09}
{Dunne} L. {et~al.}, 2009, \mnras, 394, 3

\bibitem[{{Dwek} {et~al}\mbox{.}(1998){Dwek}, {Arendt}, {Hauser}, {Fixsen},
  {Kelsall}, {Leisawitz}, {Pei}, {Wright}, {Mather}, {Moseley}, {Odegard},
  {Shafer}, {Silverberg}, \& {Weiland}}]{dwek98}
{Dwek} E. {et~al.}, 1998, \apj, 508, 106

\bibitem[{{Fixsen} {et~al}\mbox{.}(1998){Fixsen}, {Dwek}, {Mather}, {Bennett},
  \& {Shafer}}]{fixsen98}
{Fixsen} D.~J., {Dwek} E., {Mather} J.~C., {Bennett} C.~L., {Shafer} R.~A.,
  1998, \apj, 508, 123

\bibitem[{{Gao} {et~al}\mbox{.}(2004){Gao}, {White}, {Jenkins}, {Stoehr}, \&
  {Springel}}]{gao04}
{Gao} L., {White} S.~D.~M., {Jenkins} A., {Stoehr} F., {Springel} V., 2004,
  \mnras, 355, 819

\bibitem[{{Gispert}, {Lagache} \& {Puget}(2000){Gispert}, {Lagache}, \&
  {Puget}}]{gispert00}
{Gispert} R., {Lagache} G., {Puget} J.~L., 2000, \aap, 360, 1

\bibitem[{{Gonz{\'a}lez} {et~al}\mbox{.}(2011){Gonz{\'a}lez}, {Lacey}, {Baugh},
  \& {Frenk}}]{gonzalez11}
{Gonz{\'a}lez} J.~E., {Lacey} C.~G., {Baugh} C.~M., {Frenk} C.~S., 2011,
  \mnras, 413, 749

\bibitem[{{Gordon}(1988)}]{gordon88}
{Gordon} M.~A., 1988, \apj, 331, 509

\bibitem[{{Grossan} \& {Smoot}(2007)}]{grossan07}
{Grossan} B., {Smoot} G.~F., 2007, \aap, 474, 731

\bibitem[{{Guo} {et~al}\mbox{.}(2010){Guo}, {White}, {Li}, \&
  {Boylan-Kolchin}}]{guo10}
{Guo} Q., {White} S., {Li} C., {Boylan-Kolchin} M., 2010, \mnras, 404, 1111

\bibitem[{{Haiman} \& {Knox}(2000)}]{haiman00}
{Haiman} Z., {Knox} L., 2000, \apj, 530, 124

\bibitem[{{Hall} {et~al}\mbox{.}(2010){Hall}, {Keisler}, {Knox}, {Reichardt},
  {Ade}, {Aird}, {Benson}, {Bleem}, {Carlstrom}, {Chang}, {Cho}, {Crawford},
  {Crites}, {de Haan}, {Dobbs}, {George}, {Halverson}, {Holder}, {Holzapfel},
  {Hrubes}, {Joy}, {Lee}, {Leitch}, {Lueker}, {McMahon}, {Mehl}, {Meyer},
  {Mohr}, {Montroy}, {Padin}, {Plagge}, {Pryke}, {Ruhl}, {Schaffer}, {Shaw},
  {Shirokoff}, {Spieler}, {Stalder}, {Staniszewski}, {Stark}, {Switzer},
  {Vanderlinde}, {Vieira}, {Williamson}, \& {Zahn}}]{hall10}
{Hall} N.~R. {et~al.}, 2010, \apj, 718, 632

\bibitem[{{Hansen} {et~al}\mbox{.}(2009){Hansen}, {Sheldon}, {Wechsler}, \&
  {Koester}}]{hansen09}
{Hansen} S.~M., {Sheldon} E.~S., {Wechsler} R.~H., {Koester} B.~P., 2009, \apj,
  699, 1333

\bibitem[{{Hopkins} {et~al}\mbox{.}(2010){Hopkins}, {Younger}, {Hayward},
  {Narayanan}, \& {Hernquist}}]{hopkins10}
{Hopkins} P.~F., {Younger} J.~D., {Hayward} C.~C., {Narayanan} D., {Hernquist}
  L., 2010, \mnras, 402, 1693

\bibitem[{{Kennicutt}(1998)}]{kennicutt98}
{Kennicutt}, Jr. R.~C., 1998, \araa, 36, 189

\bibitem[{{Kere{\v s}} {et~al}\mbox{.}(2005){Kere{\v s}}, {Katz}, {Weinberg},
  \& {Dav{\'e}}}]{keres05}
{Kere{\v s}} D., {Katz} N., {Weinberg} D.~H., {Dav{\'e}} R., 2005, \mnras, 363,
  2

\bibitem[{{Knox} {et~al}\mbox{.}(2001){Knox}, {Cooray}, {Eisenstein}, \&
  {Haiman}}]{knox01}
{Knox} L., {Cooray} A., {Eisenstein} D., {Haiman} Z., 2001, \apj, 550, 7

\bibitem[{{Komatsu} {et~al}\mbox{.}(2011){Komatsu}, {Smith}, {Dunkley},
  {Bennett}, {Gold}, {Hinshaw}, {Jarosik}, {Larson}, {Nolta}, {Page},
  {Spergel}, {Halpern}, {Hill}, {Kogut}, {Limon}, {Meyer}, {Odegard}, {Tucker},
  {Weiland}, {Wollack}, \& {Wright}}]{komatsu11}
{Komatsu} E. {et~al.}, 2011, \apjs, 192, 18

\bibitem[{{Kravtsov} {et~al}\mbox{.}(2004){Kravtsov}, {Berlind}, {Wechsler},
  {Klypin}, {Gottl{\"o}ber}, {Allgood}, \& {Primack}}]{kravtsov04}
{Kravtsov} A.~V., {Berlind} A.~A., {Wechsler} R.~H., {Klypin} A.~A.,
  {Gottl{\"o}ber} S., {Allgood} B., {Primack} J.~R., 2004, \apj, 609, 35

\bibitem[{{Lagache} {et~al}\mbox{.}(2007){Lagache}, {Bavouzet},
  {Fernandez-Conde}, {Ponthieu}, {Rodet}, {Dole}, {Miville-Desch{\^e}nes}, \&
  {Puget}}]{lagache07}
{Lagache} G., {Bavouzet} N., {Fernandez-Conde} N., {Ponthieu} N., {Rodet} T.,
  {Dole} H., {Miville-Desch{\^e}nes} M.-A., {Puget} J.-L., 2007, \apjl, 665,
  L89

\bibitem[{{Lagache} \& {Puget}(2000)}]{LP00}
{Lagache} G., {Puget} J.~L., 2000, \aap, 355, 17

\bibitem[{{Lewis} \& {Bridle}(2002)}]{lewis02}
{Lewis} A., {Bridle} S., 2002, Phys. Rev. D, 66, 103511

\bibitem[{{Limber}(1954)}]{limber54}
{Limber} D.~N., 1954, \apj, 119, 655

\bibitem[{{Mathis} \& {Whiffen}(1989)}]{mathis89}
{Mathis} J.~S., {Whiffen} G., 1989, \apj, 341, 808

\bibitem[{{Matsuhara} {et~al}\mbox{.}(2000){Matsuhara}, {Kawara}, {Sato},
  {Taniguchi}, {Okuda}, {Matsumoto}, {Sofue}, {Wakamatsu}, {Cowie}, {Joseph},
  \& {Sanders}}]{Matsuhara+00}
{Matsuhara} H. {et~al.}, 2000, \aap, 361, 407

\bibitem[{{Meny} {et~al}\mbox{.}(2007){Meny}, {Gromov}, {Boudet}, {Bernard},
  {Paradis}, \& {Nayral}}]{meny07}
{Meny} C., {Gromov} V., {Boudet} N., {Bernard} J.-P., {Paradis} D., {Nayral}
  C., 2007, \aap, 468, 171

\bibitem[{{Nagai} \& {Kravtsov}(2005)}]{nagai05}
{Nagai} D., {Kravtsov} A.~V., 2005, \apj, 618, 557

\bibitem[{{Navarro}, {Frenk} \& {White}(1997){Navarro}, {Frenk}, \&
  {White}}]{navarro97}
{Navarro} J.~F., {Frenk} C.~S., {White} S.~D.~M., 1997, \apj, 490, 493

\bibitem[{{Neistein} \& {Dekel}(2008)}]{neistein08}
{Neistein} E., {Dekel} A., 2008, \mnras, 383, 615

\bibitem[{{Neistein} {et~al}\mbox{.}(2011){Neistein}, {Weinmann}, {Li}, \&
  {Boylan-Kolchin}}]{neistein11}
{Neistein} E., {Weinmann} S.~M., {Li} C., {Boylan-Kolchin} M., 2011, \mnras,
  414, 1405

\bibitem[{{Noeske} {et~al}\mbox{.}(2007){Noeske}, {Weiner}, {Faber},
  {Papovich}, {Koo}, {Somerville}, {Bundy}, {Conselice}, {Newman},
  {Schiminovich}, {Le Floc'h}, {Coil}, {Rieke}, {Lotz}, {Primack}, {Barmby},
  {Cooper}, {Davis}, {Ellis}, {Fazio}, {Guhathakurta}, {Huang}, {Kassin},
  {Martin}, {Phillips}, {Rich}, {Small}, {Willmer}, \& {Wilson}}]{noeske07}
{Noeske} K.~G. {et~al.}, 2007, \apjl, 660, L43

\bibitem[{{Oliver} {et~al}\mbox{.}(2010{\natexlab{a}}){Oliver}, {Frost},
  {Farrah}, {Gonzalez-Solares}, {Shupe}, {Henriques}, {Roseboom},
  {Alfonso-Luis}, {Babbedge}, {Frayer}, {Lencz}, {Lonsdale}, {Masci},
  {Padgett}, {Polletta}, {Rowan-Robinson}, {Siana}, {Smith}, {Surace}, \&
  {Vaccari}}]{oliver10a}
{Oliver} S. {et~al.}, 2010{\natexlab{a}}, \mnras, 405, 2279

\bibitem[{{Oliver} {et~al}\mbox{.}(2010{\natexlab{b}}){Oliver}, {Wang},
  {Smith}, {Altieri}, {Amblard}, {Arumugam}, {Auld}, {Aussel}, {Babbedge},
  {Blain}, {Bock}, {Boselli}, {Buat}, {Burgarella}, {Castro-Rodr{\'{\i}}guez},
  {Cava}, {Chanial}, {Clements}, {Conley}, {Conversi}, {Cooray}, {Dowell},
  {Dwek}, {Eales}, {Elbaz}, {Fox}, {Franceschini}, {Gear}, {Glenn}, {Griffin},
  {Halpern}, {Hatziminaoglou}, {Ibar}, {Isaak}, {Ivison}, {Lagache},
  {Levenson}, {Lu}, {Madden}, {Maffei}, {Mainetti}, {Marchetti},
  {Mitchell-Wynne}, {Mortier}, {Nguyen}, {O'Halloran}, {Omont}, {Page},
  {Panuzzo}, {Papageorgiou}, {Pearson}, {P{\'e}rez-Fournon}, {Pohlen},
  {Rawlings}, {Raymond}, {Rigopoulou}, {Rizzo}, {Roseboom}, {Rowan-Robinson},
  {S{\'a}nchez Portal}, {Savage}, {Schulz}, {Scott}, {Seymour}, {Shupe},
  {Stevens}, {Symeonidis}, {Trichas}, {Tugwell}, {Vaccari}, {Valiante},
  {Valtchanov}, {Vieira}, {Vigroux}, {Ward}, {Wright}, {Xu}, \&
  {Zemcov}}]{oliver10b}
{Oliver} S.~J. {et~al.}, 2010{\natexlab{b}}, \aap, 518, L21+

\bibitem[{{Reach} {et~al}\mbox{.}(1995){Reach}, {Dwek}, {Fixsen}, {Hewagama},
  {Mather}, {Shafer}, {Banday}, {Bennett}, {Cheng}, {Eplee}, {Leisawitz},
  {Lubin}, {Read}, {Rosen}, {Shuman}, {Smoot}, {Sodroski}, \&
  {Wright}}]{reach95}
{Reach} W.~T. {et~al.}, 1995, \apj, 451, 188

\bibitem[{{Rodighiero} {et~al}\mbox{.}(2010){Rodighiero}, {Cimatti},
  {Gruppioni}, {Popesso}, {Andreani}, {Altieri}, {Aussel}, {Berta},
  {Bongiovanni}, {Brisbin}, {Cava}, {Cepa}, {Daddi}, {Dominguez-Sanchez},
  {Elbaz}, {Fontana}, {F{\"o}rster Schreiber}, {Franceschini}, {Genzel},
  {Grazian}, {Lutz}, {Magdis}, {Magliocchetti}, {Magnelli}, {Maiolino},
  {Mancini}, {Nordon}, {Perez Garcia}, {Poglitsch}, {Santini},
  {Sanchez-Portal}, {Pozzi}, {Riguccini}, {Saintonge}, {Shao}, {Sturm},
  {Tacconi}, {Valtchanov}, {Wetzstein}, \& {Wieprecht}}]{rodighiero10}
{Rodighiero} G. {et~al.}, 2010, \aap, 518, L25+

\bibitem[{{Sanders} {et~al}\mbox{.}(1988){Sanders}, {Soifer}, {Elias},
  {Madore}, {Matthews}, {Neugebauer}, \& {Scoville}}]{Sanders+88}
{Sanders} D.~B., {Soifer} B.~T., {Elias} J.~H., {Madore} B.~F., {Matthews} K.,
  {Neugebauer} G., {Scoville} N.~Z., 1988, \apj, 325, 74

\bibitem[{{Schaerer} \& {de Barros}(2010)}]{schaerer10}
{Schaerer} D., {de Barros} S., 2010, \aap, 515, A73+

\bibitem[{{Sheth}(2005)}]{sheth05}
{Sheth} R.~K., 2005, \mnras, 364, 796

\bibitem[{{Shirokoff} {et~al}\mbox{.}(2011){Shirokoff}, {Reichardt}, {Shaw},
  {Millea}, {Ade}, {Aird}, {Benson}, {Bleem}, {Carlstrom}, {Chang}, {Cho},
  {Crawford}, {Crites}, {de Haan}, {Dobbs}, {Dudley}, {George}, {Halverson},
  {Holder}, {Holzapfel}, {Hrubes}, {Joy}, {Keisler}, {Knox}, {Lee}, {Leitch},
  {Lueker}, {Luong-Van}, {McMahon}, {Mehl}, {Meyer}, {Mohr}, {Montroy},
  {Padin}, {Plagge}, {Pryke}, {Ruhl}, {Schaffer}, {Spieler}, {Staniszewski},
  {Stark}, {Story}, {Vanderlinde}, {Vieira}, {Williamson}, \&
  {Zahn}}]{shirokoff11}
{Shirokoff} E. {et~al.}, 2011, \apj, 736, 61

\bibitem[{{Skibba} {et~al}\mbox{.}(2006){Skibba}, {Sheth}, {Connolly}, \&
  {Scranton}}]{skibba06}
{Skibba} R., {Sheth} R.~K., {Connolly} A.~J., {Scranton} R., 2006, \mnras, 369,
  68

\bibitem[{{Tinker} {et~al}\mbox{.}(2008){Tinker}, {Kravtsov}, {Klypin},
  {Abazajian}, {Warren}, {Yepes}, {Gottl{\"o}ber}, \& {Holz}}]{tinker08}
{Tinker} J., {Kravtsov} A.~V., {Klypin} A., {Abazajian} K., {Warren} M.,
  {Yepes} G., {Gottl{\"o}ber} S., {Holz} D.~E., 2008, \apj, 688, 709

\bibitem[{{Tinker}, {Wechsler} \& {Zheng}(2010){Tinker}, {Wechsler}, \&
  {Zheng}}]{tinker10a}
{Tinker} J.~L., {Wechsler} R.~H., {Zheng} Z., 2010, \apj, 709, 67

\bibitem[{{Tinker} \& {Wetzel}(2010)}]{tinker10b}
{Tinker} J.~L., {Wetzel} A.~R., 2010, \apj, 719, 88

\bibitem[{{Vale} \& {Ostriker}(2006)}]{vale06}
{Vale} A., {Ostriker} J.~P., 2006, \mnras, 371, 1173

\bibitem[{{Viero} {et~al}\mbox{.}(2009){Viero}, {Ade}, {Bock}, {Chapin},
  {Devlin}, {Griffin}, {Gundersen}, {Halpern}, {Hargrave}, {Hughes}, {Klein},
  {MacTavish}, {Marsden}, {Martin}, {Mauskopf}, {Moncelsi}, {Negrello},
  {Netterfield}, {Olmi}, {Pascale}, {Patanchon}, {Rex}, {Scott}, {Semisch},
  {Thomas}, {Truch}, {Tucker}, {Tucker}, \& {Wiebe}}]{viero09}
{Viero} M.~P. {et~al.}, 2009, \apj, 707, 1766

\bibitem[{{Wang} {et~al}\mbox{.}(2006){Wang}, {Li}, {Kauffmann}, \& {De
  Lucia}}]{wang06}
{Wang} L., {Li} C., {Kauffmann} G., {De Lucia} G., 2006, \mnras, 371, 537

\bibitem[{{Weinmann}, {Neistein} \& {Dekel}(2011){Weinmann}, {Neistein}, \&
  {Dekel}}]{weinmann11}
{Weinmann} S.~M., {Neistein} E., {Dekel} A., 2011, ArXiv e-prints

\bibitem[{{Wetzel} \& {White}(2010)}]{wetzel10}
{Wetzel} A.~R., {White} M., 2010, \mnras, 403, 1072

\bibitem[{{Yabe} {et~al}\mbox{.}(2009){Yabe}, {Ohta}, {Iwata}, {Sawicki},
  {Tamura}, {Akiyama}, \& {Aoki}}]{yabe09}
{Yabe} K., {Ohta} K., {Iwata} I., {Sawicki} M., {Tamura} N., {Akiyama} M.,
  {Aoki} K., 2009, \apj, 693, 507

\bibitem[{{Yang} {et~al}\mbox{.}(2005){Yang}, {Mo}, {Jing}, \& {van den
  Bosch}}]{yang05}
{Yang} X., {Mo} H.~J., {Jing} Y.~P., {van den Bosch} F.~C., 2005, \mnras, 358,
  217

\bibitem[{{Yang}, {Mo} \& {van den Bosch}(2003){Yang}, {Mo}, \& {van den
  Bosch}}]{yang03}
{Yang} X., {Mo} H.~J., {van den Bosch} F.~C., 2003, \mnras, 339, 1057

\bibitem[{{Zheng} {et~al}\mbox{.}(2005){Zheng}, {Berlind}, {Weinberg},
  {Benson}, {Baugh}, {Cole}, {Dav{\'e}}, {Frenk}, {Katz}, \& {Lacey}}]{zheng05}
{Zheng} Z. {et~al.}, 2005, \apj, 633, 791

\end{thebibliography}

\label{lastpage}
\end{document}